\documentclass[10pt]{article} 
\usepackage{amssymb, amsmath, amsthm, eucal} 
\usepackage[arrow,matrix,curve,cmtip,ps]{xy}

\newcounter{thaler}

\usepackage{tikz}
\tikzset{help lines/.style=very thin}

\renewcommand{\bar}{\overline}

\newcommand{\0}{{\bf 0}}
\renewcommand{\1}{{\bf 1}}

\newcommand{\A}{\boldsymbol{\mathcal{M}}}
\newcommand{\AG}{{\boldsymbol{\mathcal G}}} 
\newcommand{\Gri}{{\boldsymbol{\mathcal Gri}}}
\newcommand{\Gra}{{\boldsymbol{\mathcal Gra}}}
\newcommand{\B}{{\boldsymbol{\mathcal G}}}
\newcommand{\Bomega}{{\boldsymbol{\Omega}}}

\newcommand{\Bi}{{\mathcal B}} 
\newcommand{\G}{{\cal G}}  
\newcommand{\E}{{\boldsymbol{E}}}   
\newcommand{\e}{\boldsymbol{e}}
\newcommand{\f}{\boldsymbol{f}}
\newcommand{\F}{\boldsymbol{F}}   
\renewcommand{\H}{\boldsymbol{\mathcal H}}   
\newcommand{\PH}{{\mathbb P}{\boldsymbol{\mathcal H}}} 
\newcommand{\K}{\boldsymbol{K}} 
\newcommand{\M}{{\mathbf M}}   
\renewcommand{\O}{{\mathcal O}}
\newcommand{\p}{ P }     
\renewcommand{\L}{{\cal L}} 
\newcommand{\Lh}{{\cal L}_{h}} 
\newcommand{\V}{{\mathbf V}} 
\newcommand{\W}{{\mathbf W}} 

\newcommand{\R}{{\mathbb R}} 
\newcommand{\Cx}{{\mathbb C}} 
\newcommand{\N}{{\mathbb N}} 
\newcommand{\Qt}{{\mathbb H}} 

\newcommand{\C}{{\cal C}}  
\newcommand{\cc}{{\cal C}} 
\newcommand{\D}{{\cal D}}  

\newcommand{\g}{{\mathfrak g}}
\renewcommand{\p}{{\mathfrak p}}

\newcommand{\Aut}{\text{Aut}}
\newcommand{\id}{\text{id}}
\newcommand{\Aff}{\text{Aff}}

\newcommand{\Tr}{\text{Tr}}
\newcommand{\mintensor}{\otimes_{\text{min}}}
\newcommand{\maxtensor}{\otimes_{\text{max}}}

\newcommand{\tr}{\text{Tr}}

\newcommand{\cl}{\mbox{cl}}
\renewcommand{\hat}{\widehat}
\newcommand{\beqa}{\begin{eqnarray}}
\newcommand{\eeqa}{\end{eqnarray}}

\def\homega{\hat{\omega}}

\def\face{{\rm Face}}

\newtheorem{theorem}{Theorem}[]
\newtheorem{lemma}{Lemma}
\newtheorem{proposition}[theorem]{Proposition}
\newtheorem{corollary}[theorem]{Corollary}

\newtheorem{example}{Example}

\newtheorem{definition}{Definition}

\theoremstyle{remark}

\title{{\bf Post-Classical Probability Theory}} 
  \author{Howard Barnum\footnote{Department of Physics and Astronomy, University of New Mexico; {\tt hnbarnum@aol.com, hbarnum@unm.edu}} \footnote{Stellenbosch Institute for Advanced Study (STIAS), 
  Wallenberg Research Centre at Stellenbosch University, Marais Street, Stellenbosch 7600
South Africa} and Alexander
  Wilce\footnote{Department of Mathematics, Susquehanna University;
    {\tt wilce@susqu.edu}}} \date{April 13, 2012}

\begin{document}
 
\maketitle

\section{Introduction}

This chapter offers a brief introduction to what is often called the
{\em convex-operational approach} to the foundations of quantum
mechanics, and reviews selected results, mostly by ourselves and
collaborators, obtained using that approach.  Broadly speaking, the
goal of research in this vein is to locate quantum mechanics within a
very much more general, but conceptually very straightforward,
generalization of classical probability theory. The hope is that, by
regarding QM from the outside, so to say, we shall be able to
understand it more clearly. And, in fact, this proves to be the case.

The phrase ``convex-operational" deserves some comment. The approach
discussed here is ``convex" in that it takes the space of states of a
physical system to be a convex set (to accommodate the formation of
probabilistic mixtures), and draws conclusions from the geometry of this
set. It is ``operational" in its acceptance of measurements and their
outcomes as part of its the primitive conceptual apparatus, and in its
identification of states with probability weights on measurement
outcomes.  In this sense, it is conceptually very conservative,
differing from classical probability only in that it is {\em not}
assumed that all measurements can be made simultaneously.

From this starting point, one is led very naturally to a mathematical
framework for a {\em post-classical} probability theory, which, while
varying idiomatically from author to author \cite{BBLW06,
  Davies-Lewis, Edwards, Foulis-Randall74, Hardy, Holevo, Ludwig,
  Mielnik}, is more or less canonical.  About the first third of what
follows is devoted to a detailed discussion of the structure of
individual probabilistic models in this framework. Here we exhibit a
range of simple non-classical examples, many of them quite different
from {\em either} classical or quantum probabilistic models. At the
same time, we try to bring some order to this diversity, by showing
that essentially any probabilistic model can be represented in a
natural way in terms of an ordered real vector space and its dual, and
that processes operating on and between models can be reresented by
positive linear maps between these associated spaces. 

Starting in Section 3, we focus on {\em composites} of probabilistic
models, subject to a natural non-signaling constraint. As we shall
see, 
the phenomenon of {\em entanglement}, often
regarded as a hallmark of quantum mechanics, is actually a rather
generic feature of non-signaling composites of non-classical state
spaces, and thus, more a marker of non-classicality than of
``quantumness" {\em per se}. Since 
quantum information theory treats entanglement as a resource, the
question then arises of which quantum-information theoretic results
can be made to work in a more general probabilistic setting. Section 4
reviews some work in this direction, particularly the generalization
of the no-cloning and no-broadcasting theorems of
\cite{BBLW06,BBLW07}, and the analysis of teleportation and
entanglement-swapping protocols in terms of conditional states,
following \cite{BBLW08}.

If many non-classical features of QM are not so much quantum as
generically non-classical, what {\em does} single out QM? The question
of how to characterize QM in operational or probabilistic terms is a
very old one.
After many decades of hard-won partial results in this direction
(e.g., \cite{Araki, Araki-Amemiya, Birkhoff-vonNeumann, Gunson, Piron,
  Soler, Zierler}), the past decade has produced a slew of novel
derivations of finite-dimensional QM from fairly simple, transparent
and plausible, assumptions \cite{CDP,Dakic-Brukner, Hardy,
  Masanes-Mueller, Rau} (to cite just a few).  In Section 5, we
outline one of these, which recovers the Jordan structure of
finite-dimensional quantum theory from symmetry considerations; the
specific $C^{\ast}$-algebraic machinery of standard quantum mechanics
is then singled out by considerations involving the formation of
composite systems. The key tools here are a classical representation
theorem for homogeneous, self-dual cones, due to M. Koecher and
E. Vinberg \cite{Koecher, Vinberg}, and a theorem about tensor
products of Jordan algebras due to H. Hanche Olsen
\cite{Hanche-Olsen}.

Since the aim of this paper is to provide a brief and accessible
introduction to this material, we make some simplifying
assumptions. The most important is that we focus {\em entirely} on
finite-dimensional models, even though large parts of the apparatus
developed here work perfectly well (and were first developed) in an
infinite-dimensional setting.  Further assumptions will be spelled out
as we go. \\

\noindent{\em Notational conventions} Real vector spaces are indicated
generically by bold capitals $\E, \F$, etc. The space of linear
mappings $\E \rightarrow \F$ is denoted by $\L(\E,\F)$; $\E^{\ast}$
denotes the dual space of $\E$. If $\H$ is a real or complex Hilbert
space, $\Lh(\H)$ stands for the space of bounded Hermitian operators
on $\H$.  If $X$ is a set, $\R^{X}$ denotes the vector space of all
real-valued functions on $X$.

\section{Elementary probability theory, classical and otherwise}  

If $\H$ is a Hilbert space, representing a quantum-mechanical system,
then each state of that system is represented by a density operator
$\rho$. A possible measurement outcome is represented by an {\em
  effect}, i.e., a positive hermitian operator $a$ with $\0 \leq a
\leq \1$; $\Tr(\rho a)$ gives the {\em probability} that $a$ will
occur (if measured) when the state $\rho$ obtains. This probabilistic
apparatus generalizes that of classical probability theory, in that if
we fix an {\em observable}, that is, a set $\{a_1,...,a_n\}$ of
effects summing to $\1$, we can understand this as a model of a
single, discrete, classical statistical experiment, on which each
state $\rho$ defines a probability weight $p(i) := \Tr(\rho a_i)$. The
novelty here is that, in general, a pair of observables
$\{a_1,...,a_n\}$ and $\{b_1,...b_k\}$ is not {\em co-measurable}. In
classical probability theory, it is always assumed (if often tacitly)
that any pair of outcome-sets $E_1$ and $E_2$ admit a simultaneous
refinement, that is, both can be represented as partitions or
``coarse-grainings" of some third outcome-set $F$.  In
quantum-probability theory, this is not the case. Unless the operators
$a_i$ and $b_j$ all commute, there will be no third observable of
which $E_i$ are both coarse-grainings.

So, quantum probability theory foregoes the assumption of
co-measurability, which is a tenet of classical probability
theory. And, indeed, in retrospect, the latter is surely a {\em
  contingent} matter, so it is not so very radical a step to renounce
it.  It is not so much the intuitive notion of probability that is
post-classical, as the overall framework, which is in a precise sense
a generalization of the framework of the classical mathematical theory
of probability.  On the other hand, quantum probability theory
replaces the simple axiom of co-measurability with the elaborate
apparatus of the Hilbert space $\H$ and its associated space of
Hermitian operators. As a framework for an autonomous probability
calculus, this seems less than perfectly well motivated, and one can
wonder whether, and why, it is necessary. A sensible way to approach
this question is simply to drop the co-measurability assumption,
without making any special assumptions to replace it. The resulting
{\em post-classical} probability theory
is a vast, poorly explored, and rather wild region, 
within which even quantum probability theory seems rather tame. 

\subsection{Test spaces and probabilistic models}\label{subsec: models}

There are many more or less equivalent, but stylistically diverse,
ways of formulating a post-classical probability theory. The approach
we take here (due originally to C. H. Randall and D. J. Foulis
\cite{Foulis-Randall69, Foulis-Randall74}) begins with a very minimum
of raw material.

\begin{definition} A {\em test space} is a pair $(X,\A)$ where $X$ is a set of {\em outcomes} and $\A$ is a covering of $X$ by non-empty sets called {\em tests}. A {\em probablity weight} on $(X,\A)$ is a function $\alpha : X \rightarrow [0,1]$ with $\sum_{x \in E} \alpha(x) = 1$ for every $E \in \A$.  \end{definition}

The indended interpretation is that each $E \in \A$ is the set of
mutually exclusive outcomes associated with some probabilistic
experiment --- anything from rolling a die to asking a question to
making a measurement (via some well-defined procedure) of some
physical quantity.  It is permitted that distinct tests may overlap,
that is, that distinct experiments may share some outcomes. The
definition of a probability weight requires that, when this is the
case, the probability of a given outcome be independent of the
measurement used to secure it. In other words, probability weights are
{\em non-contextual}.\footnote{The formalism easily accommodates
  contextual probability assignments, however: simply define
  $\widetilde{X}$ to be the disjoint union of the test in $\A$ ---
  say, to be concrete, $\widetilde{X} = \{ (x,E) | x \in E \in
  \A\}$. In effect, each outcome of $\widetilde{X}$ consists of an
  outcome of $X$, plus a {\em record} of which test was used to secure
  it. For each test $E \in \A$, let $\widetilde{E} = \{ (x,E) | x \in
  E\}$, and let $\widetilde{\A} = \{ \widetilde{E} | E \in
  \A\}$. Probability weights on $(\widetilde{X},\widetilde{\A})$ are
  exactly what one means by {\em contextual} probability weights on
  $(X,\A)$.  There is a natural surjection $\widetilde{X} \rightarrow
  X$ that simply forgets these records; probability weights on
  $(X,\A)$ pull back along this surjection to give us weights on
  $(\widetilde{X},\widetilde{\A})$.}

It will be convenient to use the same letter, $X$, to denote the entire 
test space $(X,\A)$, as well as its outcome-set, leaving the set of tests tacit. When necessary, we'll write  $\A(X)$ for the latter.  We also write $\Bomega(X)$ for the set of all probability weights on $X$. This is a {\em convex} subset of $[0,1]^{X} \subseteq \R^{X}$, i.e., 
\[\alpha, \beta \in \Bomega(X) \ \Rightarrow \ t\alpha + (1-t)\beta \in \Bomega(X)\]
for all $0 \leq t \leq 1$.  Where $X$ is {\em locally finite}, 
meaning that every test $E \in \A(X)$ is a finite set, it is not hard to see that $\Bomega(X)$ is closed, and hence 
compact, with respect to the product topology on $[0,1]^{X}$. It follows that $\Bomega(X)$ is the closed convex hull of its extreme points. \\


\noindent{\bf Models} In constructing a model for a probabilistic
system, we may wish to single out certain probability weights as
corresponding to possible {\em states} of the system. It is reasonable
to form probability-weighted averages of such states, in order to
represent ensembles of systems in different states. It is also
reasonable to idealize the situation slightly by assuming that the
limit of a sequence of possible states should again count as a
possible state.  In the same spirit, we shall assume in what follows
that $X$ carries a Hausdorff topology, with respect to which states
are continuous. This is harmless, since we can always use the discrete
topology as a default.\footnote{A more detailed discussion of test
  spaces with topological structure can be found in \cite{Wilce05a}}
Indeed, given that our focus here is exclusively on finite-dimensional
models, it is not unreasonable to assume that $X$ is even compact.

To make all of this official:



\begin{definition} {\em A {\em probabilistic model} --- or, for purposes of this paper, just a {\em model} --- is a structure $(X,\Omega)$, where  $X$ is a 
Hausdorff test space and $\Omega$ is a pointwise-closed (hence, compact), convex set of continuous 
probability weights on $\Omega(X)$.  The extreme points of $\Omega$
are the {\em pure states} of the model.  }\end{definition}

\noindent{\em Notation:} We henceforth use capital letters $A$, $B$,
etc. to denote models, writing, e.g., $(X(A),\A(A))$ for the test 
space belonging to model $A$, and $\Omega(A)$ for
$A$'s state space.  (So technically, $A = ((X(A), \A(A)), \Omega(A))$.)




\begin{example}[Classical Models] 
{\em (a) The simplest classical models have the structure $(E,\Delta(E))$,
  where $E$ is a single test (so that $\A(E) = \{E\}$), and where and
  $\Delta(E)$ is the simplex of all probability weights thereon.  We
  might also deem ``classical" a broader set of models: those of the
  form $(E,\Omega)$ where $\Omega \subseteq \Delta(E)$ is any closed,
  convex set of probability weights sufficiently large to 
  statistically separate different outcomes\footnote{That is, given any pair of distinct outcomes, there exists
a state assigning them different probabilities.} of the single test $E$. \\

\noindent(b) A more sophisticated classical model begins with a
measurable space $S$, and identifies statistical experiments with
finite or countably infinite partitions of $S$ by measurable subsets.
The collection of all such experiments is a test space: 
let $X(S)$ be the set of non-empty measurable subsets of $S$ (say,
with the discrete topology), and let $\D(S)$ be the set of countable
partitions of $S$ into measurable subsets. We call $(X(S),\D(S))$ the
{\em Kolmogorovian} test space associated with $S$. Probability weight
on $(X(S),\D(S))$ correspond exactly to countably-additive probability
measures on $S$.\footnote{By varying $\D(S)$, we can change the
  character of the probability weights that are allowed. For example,
  if we let $\D(S)$ include just the {\em finite} measurable
  partitions of $S$, then probability weights on $\D(S)$ correspond to
  finitely additive measures on $S$. }}
\end{example}

\begin{example}[\bf Quantum Models]\label{ex: quantum models} 
{\em (a) The most basic quantum-mechanical model begins with a  complex Hilbert space $\H$.  The 
{\em quantum test space} is  $(X(\H),\M(\H))$ where the outcome space $X(\H)$ is the unit sphere of $\H$ (with its 
usual topology) and where the space $\M(\H)$ of tests is the set of unordered orthonormal bases of {\em frames} of $\H$. Every unit vector $v \in \H$ determines a probability weight $\alpha_{v}$  on $\M(\H)$, defined for all $x \in X(\H)$ by \[\alpha_{v}(x) = |\langle v, x \rangle|^2 = \Tr(P_{v} P_x),\] 
where $P_v$ and $P_x$ are the rank-one projection operators corresponding to $v$ and $x$. Accordingly, if $W$ is a density operator on $\H$ --- a positive hermitian operator of trace one, or, equivalently, a convex combination of rank-one projections --- then $\alpha_{W}(x) := \langle W x, x \rangle = \Tr(WP_x)$ defines a probability weight on $X(\H)$. If $\dim(\H) \geq 3$, then Gleason's theorem tells us that every probability weight on $X(\H)$ is of this form, but for $\dim(\H) = 2$, there are many others, which one regards as non-physical. In either case, letting $\Omega(\H)$ denote the convex set of density operators on $\H$, we obtain the {\em quantum model} $A(\H) = (X(\H),\Omega(\H))$. 

A slightly different model, which we'll call the {\em projective} quantum model, and which we denote by $A(\PH)$, replaces each outcome $x \in X(\PH)$ by the corresponding 
rank-one projection operator $P_x$; tests in $\M(\PH)$ are maximal pairwise orthogonal families of such projections. Again, states correspond to density operators via the recipe $\alpha_{W}(P_{x}) = \Tr(WP_{x})$ where 
$P_{x} \in X(\PH)$. For many purposes, the choice between $A(\H)$ and $A(\PH)$ is one of convenience. However, notice that in passing from $A(\H)$ to $A(\PH)$ we lose information about phase relations between the unit vectors representing outcomes of $X(\H)$, which are important in describing sequential experiments. We won't pursue this here. The paper \cite{Wright} contains 
some relevant discussion. \\

\noindent (b) A more sophisticated quantum model might begin with a
$W^{\ast}$-algebra $\cal A$, and take for $\A$, the collection of all
(say, finite) sets of projections summing to the identity in $\cal
A$. If $\A$ has no $I_2$ summand, the Christensen-Yeadon extension of
Gleason's theorem \cite{Dvurecenskij} identifies the probability
weights on $\A$ with states on $\cal A$. Again, if there are $I_2$
factors (copies of $M_2(\Cx)$), then one must explicitly 
limit the states to the quantum-mechanical ones.} \end{example}

By the {\em dimension} of a model $A$, we mean the dimension of the span of $\Omega(A)$ in $\R^{X(A)}$. Of course, this will generally be infinite. However, as mentioned in the introduction, 
our focus in this paper is on finite-dimensional models. Indeed, making this official, we assume from this point forward that {\bf \emph{all models are finite-dimensional}.} In particular, all 
{\em quantum} models $A(\H)$ and $A(\PH)$ involve only finite-dimensional Hilbert spaces $\H$. 

If we let $\V(\Omega)$ denote the span of $\Omega$ in $\R^{X(A)}$, we can map $X(A)$ into $\V(A)^{\ast}$ by evaluation. That is, for each outcome $x \in X(A)$, there is a canonical 
evaluation functional $\hat{x} : \V(A) \rightarrow \R$ given by $\hat{x}(\alpha) = \alpha(x)$. 
It may happen that, for some sequence $x_i$ of outcomes, $\hat{\alpha}(x_i) \rightarrow a \in \V(A)^{\ast}$. Let us say that $A$ is {\em outcome-closed} iff every such limit again corresponds to an outcome in $X(A)$, i.e, that there  exists some $x \in X(A)$ with $a = \hat{x}$.  Where $X(A)$ is compact in its native topology --- which, in finite dimensional examples, it very often is --- this condition is automatically satisfied. We make it another standing 
assumption that {\bf \emph{all models are outcome-closed.}}\\


\noindent{\bf Dispersion-Free States and Distinguishability}  One very striking difference between classical and quantum models has to do with the existence of (globally) {\em dispersion-free}, that is, zero-or-one valued, states. In both of 
the classical models considered above, all pure states are dispersion-free. Quantum models, in contrast, have  {\em no} dispersion-free state: a pure quantum state sill makes only uncertain predictions about the results of most measurements. 

\begin{definition} {\em A set $\Omega$ of probability weights on a test space $X$ is {\em unital} iff, for every $x \in X$, there exists at least one $\alpha \in \Omega$ with $\alpha(x) = 1$. 
If there is a {\em unique} such state, we say that $\Omega$ is {\em sharp}. We say that a model $A$ is 
unital or sharp if its state space $\Omega(A)$ is a unital, respectively sharp, set of probability weights 
on the test space $X(A)$. } \end{definition}

Like the classical examples, the quantum quantum models $A(\H)$ and $A(\PH)$ are sharp;
indeed, the unique state $\alpha$ assigning probability one to a given outcome $x \in X(\H)$, or 
to the corresponding outcome $P_x \in X(\PH)$, is is the one corresponding to the density operator $P_x$.   

\begin{definition}{\em A set $\Omega$ of probability weights on a test space $X$ {\em separates outcomes}, or is {\em separating}, iff, for all outcomes $x, y \in X$, $\alpha(x) = \alpha(y)$ for all $\alpha \in \Omega$ implies $x = y$.  A model $A$ is {\em separated} iff $\Omega(A)$ separates outcomes of $X(A)$.}\end{definition}

The state space of a standard quantum model $A(\H)$ is not separating; that of the corresponding projective quantum model $A(\PH)$ {\em is} separating. As this 
example illustrates, given a non-separated model $A$, one {\em can} always replace $X(A)$ by an obvious quotient test space, in which probabilistically indistinguishable outcomes are identified, to obtain a separated model having the same sates. One may or may not {\em wish} to do so.

A {\em partition space} is a test space that is isomorphic\footnote{An isomorphism of test spaces is  a bijection from outcomes to outcomes, preserving tests in both directions.} to a sub-test space of $\D(S)$ for some set $S$. Any such space supports a state-separating set of dispersion-free probability weights, 
namely, the point-masses associated with the points of $S$. The following is straightforward:

\begin{lemma} If test space has a unital, separating set of dispersion-free states, then it is a partition test space. If it has a sharp set of unital, DF states, then it is classical.\end{lemma}

In anticipation of later results, we'll write $x \perp y$ to mean that
outcomes $x, y \in X(A)$ are {\em distinguishable} by means of some
test $E \in X(\A)$ --- that is, that $x, y \in E$ and $x \not = y$. At
present, there is no linear structure in view, let alone an inner
product, so the notation is only suggestive. Later, we'll see that one
can often embed $X$ in an inner product space in such a way that the
notation can be taken literally.

It will also be useful to introduce the following notion of distinguishability for {\em states}.

\begin{definition}{\em Two states, $\alpha, \beta \in \Omega(A)$ are {\em sharply distinguishable} iff 
there exist outcomes $x, y \in X(A)$ with $x \perp y$ such that
$\alpha(x) = \beta(y) = 1$. More generally, states
$\alpha_1,...,\alpha_n$ are {\em jointly} sharply distinguishable iff
there exists a test $E \in \A(A)$ and outcomes $x_1,...,x_n \in E$
with $\alpha_i(x_j) = \delta_{i,j}$.} \end{definition}

The idea is that, if the system is known to be in one of the states
$\alpha_1,...,\alpha_n$, then by performing the measurement $E$ we
will learn --- with probability one -- which of these states was the
actual one.\footnote{A weaker notion would require only that
  $\alpha_i(x_i) > 0 = \alpha_i(x_j)$ for each $i,j$, so that with
  {\em some} non-zero probability we obtain either $x$ or $y$, and
  thus learn which state was actual. Notice, too, that the condition
  of joint sharp distinguishability is a priori much stronger than
  pairwise sharp distinguishability.}


\subsection{Further Examples} 

Classical and quantum examples hardly exhaust the possibilities, of
course: the whole point of the present framework is to provide us with
a maximum of flexibility in constructing {\em ac hoc} models.

\begin{example}[The Square Bit] {\em The very simplest non-classical model starts with a test space $X$ be a test space containing just two tests  $E = \{x,x'\}$ and $F = \{y,y'\}$, each having 
two outcomes --- as, say, two coins, or a stern-Gerlach apparatus with
two angular settings. The convex set $\Omega(X)$ of all probability
weights on $X$ is affinely isomorphic to the unit square, under the
mapping $\alpha \mapsto (\alpha(x),\alpha(y))$. The model $(X,\Omega)$
has, accordingly, been called the {\em square bit} \cite{Entropy}. As
$\Omega(X)$ is not a simplex, this model is not entirely classical. On
the other hand, as its pure states are all dispersion-free, it is very
far from being ``quantum".}
\end{example}

\noindent{\bf Greechie Diagrams} A useful graphical device for
representing small test spaces (those involving only a few outcomes)
is to represent each outcome as a dot, and to join outcomes belonging
to a test by a straight line or other smooth arc, with arcs
corresponding to distinct tetst intersecting, if at all, at a sharp
angle, so as to be easily distinguished.  Such a representation (first
used in the quantum-logical literature) is called a {\em Greechie
  diagram} \cite{Greechie}.  For example, we might represent a
three-outcome classical test by the diagram in Figure 2 (a), and the
square-bit test space by that in Figure 2 (b). The test space pictured
in (c), with two three-outcome tests (the top and bottom rows) and
three two-outcome tests (the vertical lines), makes the point that a
test space need not have any states at all.
{\small 
\[
\begin{array}{ccccc}
{\begin{tikzpicture} 
\draw (0,0) node {$\bullet$} (1,0) node {$\bullet$} (2,0) node{$\bullet$}; 
\draw (0,0) -- (1,0) -- (2,0);
\end{tikzpicture}}
& & 
{\begin{tikzpicture}
\draw (5,.5) node {$\bullet$} (6,.5) node {$\bullet$} (5,-.5) node{$\bullet$} (6,-.5) node {$\bullet$}; 
\draw (5,.5) -- (5,-.5); \draw (6,.5) --(6,-.5);
\end{tikzpicture}}
& & 
{\begin{tikzpicture}
\draw (9,.5) node {$\bullet$} (10,.5) node {$\bullet$} (11,.5)  node {$\bullet$};
\draw (9,-.5) node{$\bullet$} (10,-.5) node {$\bullet$} (11,-.5) node {$\bullet$}; 
\draw (9,.5) -- (10,.5) -- (11,.5) -- (11,-.5) -- (10,-.5) -- (9,-.5) -- (9,.5); 
\draw (10,.5) -- (10,-.5);
\end{tikzpicture}}\\
\mbox{(a)} & & \mbox{(b)} & & \mbox{(c)} 
\end{array}\] 
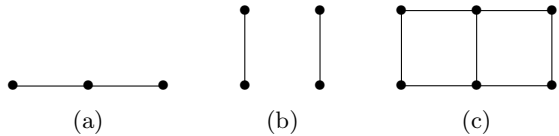
\begin{center}{\sc Figure 1: Various Greechie diagrams}
\end{center}
}

The following whimsical example (due to D. J. Foulis) is useful as an antidote to several too-comfortable intuitions.

\begin{example}[The Firefly Box] {\em  Suppose a sealed
triangular box is divided into three interior chambers, as in the top-down view in Figure 2(a), below.
The walls of the box are translucent, while the top, the bottom, and
the interior partitions are opaque. In the box is a firefly, free to
move about between the chambers (for which purpose, the interior
partitions contain small tunnels). Viewed from one side, we might
see the firefly flashing in chamber $a$ or chamber $b$, or we might
see nothing -- the firefly might not be flashing, or might be in
chamber $c$. Thus, we have three experiments, corresponding to the
three walls of the box: $\{a,x,b\}$, $\{b,y,c\}$ and $\{c,z,a\}$,
where $x, y$ and $z$ are the (distinct) ``no-light" outcomes
associated with each experiment. The resulting test space ${\mathfrak A}
= \{\{a,x,b \},\{b,y,c\},\{c,z,a\}\}$ has the Greechie diagram pictured in Figure 2(b) below.
{\small
\[ 
\begin{array}{ccccc} 
{\begin{tikzpicture} 
\draw (-30:.7) node {$b$} (90:.7) node {$c$} (210:.7) node{$a$};
\draw[double] (-30:1.5) -- (90:1.5) -- (210:1.5) -- (-30:1.5); 
\draw[thick] (-90:.76) -- (0,0) -- (150:.75); \draw[thick] (0,0) --  (30:.75); 
\end{tikzpicture}}
& & 
{\begin{tikzpicture} 
\draw (-30:1.5) node {$\bullet$} (90:1.5) node {$\bullet$} (210:1.5) node{$\bullet$};
\draw (-22:1.7) node {$b$} (90:1.7) node {$c$} (202:1.7) node{$a$};
\draw (-30:1.5) -- (90:1.5) -- (210:1.5) -- (-30:1.5); 
\draw (-90:.76) node {$\bullet$} (150:.75) node {$\bullet$} (30:.75) node {$\bullet$};
\draw (-90:.55) node {$x$} (150:.89) node {$y$} (30:.89) node {$z$};
\end{tikzpicture}}
& & 
\begin{tikzpicture}
\draw (0,0) -- (1,-1.5) -- (0,2) -- (0,0); \draw (-.2,-.2) node {$\delta$};
\draw[dashed] (0,0) -- (2,0) -- (0,2); \draw (-.2,2.2) node {$\gamma$};
\draw (1,-1.5) -- (2,0); \draw (2.2,-.2) node {$\beta$}; \draw (.7,-1.3) node {$\alpha$}; 
\draw (1,-1.5) -- (1.5,.25) -- (2,0) -- (0,2);
\draw (1.5,.25) -- (0,2); 
\draw (1.3,.25) node {$\epsilon$}; 
\end{tikzpicture}\\
(a) & \hspace{.2in} & (b) & & (c) \end{array}
 \] 
 \begin{center}
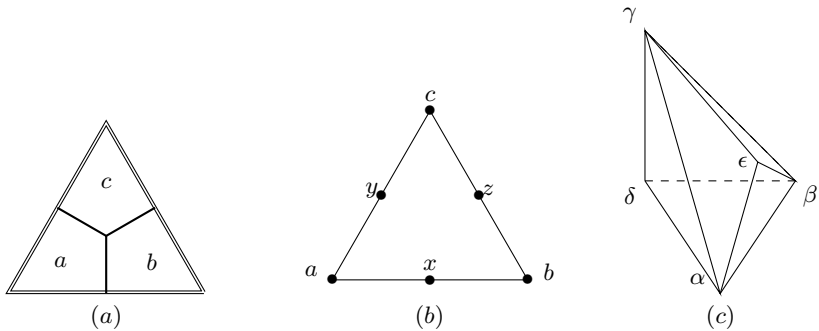
 {\sc Figure 2: The Firefly Box} 
 \end{center}
 }

We can identify several pure states on this test space with concete situations involving the 
location, and the internal state (lit or unlit) of the firefly. 
For example,
\[\alpha(a) = \alpha(z) = 1; \ \alpha(b) = \alpha(c) = \alpha(x) = \alpha(y) = 0\]
corresponds to the firefly's flashing in chamber $a$. We can define similar states 
$\beta$ and $\gamma$ corresponding to chambers $b$ and $c$. All of these states are 
dispersion-free. A fourth dispersion-free pure state, $\delta$, assigns 
probability $1$ to the outcomes $x, y$ and $z$. This corresponds to the firefly not flashing. 
These four dispersion-free states separate outcomes separate the six outcomes, and thus allow us, 
by Lemma 1, to represent the firefly box as a partition test space over a classical state space. 
{\em However}, there is  also a fifth, {\em non}-dispersion 
free pure state, $\epsilon$, given by 
\[\epsilon(a) = \epsilon(b) = \epsilon(c) = 1/2; \ \epsilon(x) = \epsilon(y) = \epsilon(z) = 0.\]
This last state is difficult to interpret in any way but to imagine that the firefly 
{\em responds} to being observed through a given window by entering (with equal probability) one of 
the two corresponding chambers. Since any state on this test space is determined by its values at the 
outcomes $x$, $y$ and $z$, the convex set of all probablilty weights for the firefly box 
is a non-simplicial set in $\R^{3}$: the pure states $\alpha,\beta$ and $\gamma$ 
correspond to the standard basis vectors $(1,0,0), (0,1,0)$ and $(0,0,1)$, $\delta$ corresponds 
to the origin, and $\epsilon$, to the vector $1/2(1,1,1)$. Thus, $\Omega$ is 
affinely isomorpic to a triangular diprism, as pictured in Figure 2 (c).  
}\end{example}

\begin{example}[Grids and Graphs] {\em Let $E$ be a finite set --- for definiteness, say $\{0,1,...,n-1\}$, with $n \geq 2$. We define two test spaces associated with $E$: 
\begin{itemize}
\item[(a)] The {\em grid test space}, $\Gri(E)$, consists of all rows and columns of the $n \times n$ array $E \times E$, that is, all sets of the form $\{x\} \times E$ or $E \times \{y\}$. 
\item[(b)] The {\em graph test space}, $\Gra(E)$ consists  of the {\em graphs} of permutations $f : E \rightarrow E$, that is, subsets of $E \times E$ of the form $\{(i,f(i)) | i \in E\}$. 
\end{itemize} 
Both of these test spaces have outcome-set $X = E \times E$, so a state on either test space can be regarded as an $n \times n$ real matrix with non-negative entries. In the case of $\Gri(E)$, 
these entries must sum to unity along  each row and column; that is, the states on $\Gri(E)$ are exactly the {\em doubly stochastic matrices}. By the Birkhoff-von Neumann theorem, these 
all arise as convex combinations of permutation matrices --- that is, of the dispersion-free states corresponding to elements of $\Gra(E)$. Similarly, one can show that, for $n \geq 3$, every state of 
$\Gra(E)$ is an average of  {\em row states}, $\alpha^{k}$, given by $\alpha^{k}(i,j) = \delta_{i,k}$ and {\em column states} $\alpha_{k}$, given by $\alpha_k(i,j) = \delta_{k,j}$.

Every pair of pure states on either $\Gri(E)$ or $\Gra(E)$ is distinguishable by a test in that space. 
Nevertheless, neither state space is a simplex for $n \geq 3$. The space of doubly-stochastic matrices has $n!$ pure states, which, for $n \geq 4$, exceeds the $n^2 + 1$ states permissible for a simplex in $\R^{n^2}$. For  $n \geq 3$, $\B(E)$ has only $2n$ pure states; however, the maximally mixed state $\alpha(i,j) \equiv 1/n$, can be represented as a uniform average over just the row states, or over just the column states; similarly, on $\Gri(E)$, it can be represented as a uniform average over any set of permutations the graphs of which partition $E \times E$. By a curious coincidence, the test spaces $\Gri(3)$ and $\Gra(3)$ are isomorphic, so the state space of 
$\Gri(3)$ is isomorphic to that of $\Gra(3)$, and again, not a simplex.}\end{example}

\noindent{\em Remark:} We've seen that a variety of convex geometries can arise more or less naturally as the (full) state spaces of test spaces. A natural question is whether {\em every} possible convex geometry 
arises in this way. A theorem of F. Shultz \cite{Shultz} shows that in fact, every compact convex set can be represented as the space of probability measures on an orthomodular lattice. The set of 
decompositions of the unit element in such a lattice is a test space, the probability weights on which correspond precisely to the probability measures on the lattice. Thus, Shultz' theorem implies 
that every compact convex set can be realized as the full state space of a test space. \\

\noindent{\bf Models from Symmetry} A {\em symmetry} of a test space
$X$ is a bijection $g : X \rightarrow X$ such that both $g$ and
$g^{-1}$ preserve tests --- in other words, such that for all $E
\subseteq X$, we have $gE \in \A(X)$ iff $E \in \A(X)$. (In other
words, it is an isomorphism from the test space $X$ to itself.) The
set of all symmetries of $X$ is evidently a group, which we'll denote
by $G(X)$. There is a natural dual action of $G(X)$ on probability
weights on $X$, given by $g \alpha := \alpha \circ g^{-1}$; a symmetry
of a model $A = (X,\Omega)$ is a symmetry of $A$ that also preserves
$\Omega$. Again, the symmetries of a model form a group, $G(A) \leq
G(X(A))$.

Both classical and quantum test spaces are marked by very strong symmetry properties. In particular, the symmetry group of either kind of system acts transitively on pure states, and also 
on the set of tests; moreover, any permutation of the outcomes of any given test can be implemented by a symmetry of the entire system. (This is more or less trivial in the case of a classical system; for a quantum system, it amounts to the observation that any permutation of an 
orthonormal basis for a Hilbert space $\H$ extends to a unitary operator on $\H$.) In contrast, no symmetry 
of the ``firefly box" test space of Example 4 will exchange one of the outcomes $a,b,c$ with one of $x,y,z$, 
since each of the former belongs to two tests, while each of the latter belongs only to one.

\begin{definition}{\em Let $G$ be a group acting by symmetries on a test space $X$. We say $X$ is {\em symmetric} under $G$, or {\em $G$-symmetric}, iff $G$ acts transitively on $\A(X)$, and the stabilizer $G_{E}$ of a test 
$E \in \M(A)$ acts transitively on $E$. If $X$ is $G$-symmetric and $G_{E}$ acts doubly transitively on $E$, then 
$X$ is {\em $2$-symmetric} under $G$. If $G_{E}$ acts as the full permutation group of $E$, we say that $X$ is 
{\em fully} $G$-symmetric.} \end{definition}

 
In fact, test spaces with these symmetry properties can be constructed very naturally \cite{Wilce11}. Suppose one has a simple measuring device, which can be applied to a system of some sort to produce outcomes in a set $E$. One might be able to apply this device {\em in different ways} --- for example, by changing the orientation of the apparatus with respect to the system, or by adjusting some controllable physical parameters associated with the system. This suggests that we might be able to build a larger family of experiments --- a test space, in other words --- starting with the basic measurement $E$, and adding parameters that keep track of the various ways in which we might deploy it. 
In many cases, there will be a group $G$ of ``physical symmetries"
acting on these parameters, and we can often reconstruct the desired
test space simply from a knowledge of this group and its relationship
to the test $E$. Specifically, there will be some subgroup $H$ of $G$
that acts to permute the outcomes of $E$.  Let us suppose that $H$
acts transitively on $E$, so that, for any reference outcome $x_o \in
E$, every other outcome $x \in E$ has the form $hx_o$ for some $h \in
H$. If we let $K$ be any subgroup of $G$ such that $K \cap H =
H_{x_o}$, where $H_{x_o}$ is the stabilizer in $H$ of a chosen
reference outcome $x_o \in E$, and set $X = G/K$. Then there is a
well-defined canonical $H$-equivariant injection $j : E \rightarrow X$
given by $j(x) = hK$ where $x = h x_o$.  Let us identify $E$ with
its image under $j$, so that $E \subseteq X$, let $\AG$ be the orbit
of $E$ under $G$, i.e.,
\[\AG := \{ gE | g \in G\}.\] 
The test space $(X,\AG)$ will automatically be symmetric, and will be $2$-symmetric or fully symmetric under $G$ as 
$H$ acts doubly or fully transitively on $E$.  We obtain a $G$-symmetric {\em model} by choosing any 
$G$-invariant, closed, convex set of probability weights on $X$. 

  
  The choice of the group $K$ extending the stabilizer $H_o$ has a large effect on the 
combinatorial structure of $(X,\AG)$. For example, if $K = H_{o}$, then $\A$ is a semi-classical test space consisting of disjoint copies of $E$; in general, a larger choice of $K$ will enforce 
non-trivial intersections among the tests $gE$ with $g \in G$. 

\begin{example} {\em As an illustration of this construction, let $E = \{0,1,...,n-1\}$, and let $U$ be the group of all unitary $n \times n$ matrices, acting in the usual way on $\H = \Cx^{E}$. Let $H \leq U$ be the subgroup consisting of permutation matrices, and $K$, the group of unitaries fixing $\e_0$, the column vector corresponding to $0 \in E$.  Then $K \cap H$ is exactly the set of permutation matrices corresponding 
to permutations fixing $0$, i.e., $K \cap H = H_{0}$. Now $X = G/K$ is the (projective) unit sphere of $\H$, and $\A$ is the set of (projective) frames of $\H$.  For another 
example, let $H$ be the full permutation group $S(E)$ of $E$ and set $G = S(E) \times S(E)$. Embedding $H$ in $G$ by $h \mapsto (h,e)$, the construction above produces the ``grid" test space $\Gri(E)$ of Example 6. Using instead the diagonal embedding $h \mapsto (h,h)$ yields the ``graph" test space $\Gra(E)$.} \end{example}

\subsection{Models Linearized} 

In many situations, the outcomes of a test space are naturally represented as elements of a vector space. This 
is obviously the case for the quantum-mechanical examples discussed above, where outcomes are directly identified 
with unit vectors in $\H$ or with rank-one projections in $\L(\H)$. One can also formulate classical probability theory in this way, by considering the space of random 
variables associated with a given measurable space, and identifying measurement outcomes (that is, measurable sets) with the corresponding indicator random variables. 

In fact, subject to some fairly mild restrictions, such a representation is always available. The idea will be to construct, for each such a model $A = (X,\Omega)$, a real vector space 
$\E(A)$, and an embedding of $X \rightarrow \E(A)$, in such a way that states in $\Omega$ extend uniquely to linear functionals on $\E(A)$.  In fact, $\E(A)$ will be an {\em ordered} real vector space, so we pause briefly to review this notion (for further details, see \cite{Alfsen-Shultz}). \\

\noindent{\bf Ordered Linear Spaces} By a {\em cone} in a real vector
space $\E$, we mean a convex subset closed under multiplication by
non-negative scalars, and satisfying $K \cap -K = \{0\}$. $K$ is {\em
  generating} iff it spans $\E$.  it spans $\E$. An {\em ordered
  linear space} is a real vector space $\E$, equipped with a closed,
generating cone $\E_+$.  Such a cone determines a (partial) ordering,
invariant under translation and under positive scalar multiplication,
on $\E$, namely $a \leq b$ iff $b - a \in \E_+$.\footnote{Some authors 
define ordered linear spaces without requiring that the positive cone be
  generating. 
  For our purposes, the present definition is more useful. 
  }  
Noticing that $a \geq 0$ iff $a
\in \E_+$, we refer to $\E_+$ as the {\em positive cone} of $\E$.

The basic example is the space $\R^{X}$ of all real-valued functions on a set $X$, ordered pointwise. Thus, \[(\R^X)_+ = \{ f \in \R^{X} \ | \  f(x) \geq 0 \ \forall x \in X\}.\] 
Another example, central to our concerns here, is the space $\Lh(\H)$ of  bounded {\em hermitian} operators on a Hilbert space $\H$ (over either $\R$ or $\C)$. This space 
has a standard ordering, induced by the cone $\L_{+}(\H)$ of positive semi-definite operators --- that is, $a \in \L_{+}(\H)$ iff $\langle a x, x \rangle \geq 0$ for all vectors $x \in \H$. 
More generally, the real vector space of self-adjoint elements of a $C^{\ast}$-algebra $\mathcal A$ is ordered by the cone of elements of the form 
$aa^{\ast}$, $a \in {\mathcal A}$.

If $\E$ and $\F$ are ordered linear spaces, a linear mapping $f : \E \rightarrow \F$ is {\em positive} iff $f(\E_+) \subseteq \F_{+}$, i.e, $f(a) \geq 0$ whenever $a \geq 0$. An {\em order-isomorphism} between $\E$ and $\F$ is a 
positive, invertible linear mapping having a positive inverse. We'll denote 
the set of postive linear mappings $\E \rightarrow \F$ by $\L(\E,\F)$. This is a cone in the space $\L(\E,\F)$. 
As a special case, the dual space of an ordered vector space $\E$ has a natural {\em dual cone}, 
$\E^{\ast}_+ = \L_{+}(\E,\R)$. In our present finite-dimensional setting, this is generating, so $\E^{\ast}$ becomes 
an ordered vector space in a natural way.  \\

\noindent{\bf Order-unit spaces} An {\em order unit} in an ordered linear space $\E$ is an element $u \in \E_+$ such that, for every $a \in E$, there exists some $n \in \N$ with $a \leq nu$. When $\E$ is finite-dimensional, 
this is equivalent to asking that $\alpha(u) > 0$ for every  non-zero$\alpha \in \E^{\ast}_+$, which can always be arranged. (In particular, a finite-dimensional ordered linear space always {\em has} an order-unit.) An {\em order-unit space} is an ordered linear space equipped with a distinguished order-unit.  The key example to bear in mind is the space $\Lh(\H)$, ordered as described above, and with the identity operator as order-unit.

An order unit space already provides enough structure to support
probabilistic ideas. A {\em state} on an order-unit space $\E$ is a
linear functional $\alpha \in \E^{\ast}$ with $\alpha(u) = 1$. An {\em
  effect} in $\E$ is a positive element $a$ with $a \leq u$, so that
$0 \leq \alpha(a) \leq 1$ for every state $\alpha$.  A discrete {\em
  observable} on $\E$ is a finite set $E = \{a_1,...,a_{k}\}$ of
non-zero effects with $a_1 + \cdots + a_k = u$; evidently, any state
on $\E$ restricts to a probability weight on every observable on
$\E$. Thus, the observables form a test space, the outcomes of which
are just the non-zero effects in $\E_+$. In the special case where $\E
= L_h(\H)$, the space of Hermitian operators on a Hilbert space $\H$,
an effect is a positive opertor $a$ with $0 \leq a \leq \1$; all
states have the form $\alpha(a) = \Tr(Wa)$ where $W$ is a density
operator on $\H$, and an observable is essentially a (discrete)
positive-operator valued measure.

The set of all (normalized) states on an order-unit space $\E$ is the latter's {\em state space}. This is 
always a compact convex set. Conversely, if $\Omega$ is any compact convex subset of any finite-dimensional real vector space, let $\Aff(\Omega)$ denote the space of bounded affine (that is, convex-combination preserving) real-valued functionals 
$f : \Omega \rightarrow \R$, ordered pointwise. The constant 
functional $u(\alpha) \equiv 1$ serves as an order unit. One can show that $\Omega$ (embedded in $\Aff(\Omega)^{\ast}$ by evaluation) is exactly $\Aff(\Omega)$'s state space. Moreover, if $T : \Omega \rightarrow \W_{+}$ is any affine mapping of $\Omega$ into the positive cone of a (finite-dimensional) ordered linear space $\W$, then $T$ extends uniquely to a positive linear mapping $ T : \E(\Omega)^{\ast} \rightarrow \W$. \\

\noindent{\bf The linear hull of a model} Any probabilistic model can
be interpreted, in a canonical way, in terms of an order-unit space
with a distinguished family of observables.  Let $A = (X,\Omega)$ be a
probilistic model. Every outcome $x \in X(A)$ determines an affine
functional $\hat{x} : \Omega \rightarrow \R$ by evaluation:
$\hat{x}(\alpha) = \alpha(x)$ for all $\alpha \in \Omega$.

\begin{definition} If $A = (X,\Omega)$ is a model, write $\E(A)$ for the span of $X$ in $\R^{\Omega}$, ordered by the closure of the cone consisting of linear combinations with non-negative coefficients of evaluation functionals $\hat{x}$, $x \in X(A)$:
\[\E(A)_{+} = \cl\left ( \left \{ \ \sum_{i} t_i \hat{x}_i \ | x_i \in X, \ t_i \geq 0 \ \right \} \right ) .\]
\end{definition}

Letting $u \in \R^{\Omega}$ denote the constant function $u(\alpha)
\equiv 1$, we see that $\sum_{x \in E} \hat{x}$, where $E$ is any test
in $\A(A)$. Hence, $u$ belongs to $\E_{+}$, where it functions as an
order-unit. The order-unit space $(\E(A),u)$, togther with the
embedding $X(A) \rightarrow \E(A)$, is called the {\em linear hull} of the
model $A$. Every test $E \in \A(A)$ can now be regarded as a discrete
observable on $\E(A)$.  Notice that the cone $\E(A)_+$ may well be
smaller than the cone $\{ \ a \in \E \ | \ a(\alpha) \geq 0 \ \forall
\alpha \in \Omega \ \}$ inherited from $\Aff(\Omega(A))_{+}$, and
that, unlike the latter, it depends on the choice of $X(A)$.

\begin{example} {\em In the case of a quantum model $A = (X(\H),\A(\H))$ of Example 2,  the space $\E(A)$ --- or, 
as we'll denote it below, $\E(\H)$ --- can be identified with the
order-unit space $\Lh(\H)$ of Hermitian operators on $\H$, ordered by
the usual cone, with $u$ the identity operator.}\end{example}

There is a canonical embedding of $\Omega(A)$ in $\Aff(\Omega)^{\ast}$, taking each state $\alpha \in \Omega(A)$ with the corresponding evaluation functional $f \mapsto f(\alpha)$, $f \in \Aff(\Omega)$. Let $\V(A)$ denote the span of $\Omega(A)$ in $\Aff(\Omega)^{\ast}$, ordered by the cone $\V_+(A)$ generated by $\Omega(A)$.  Since $\E(A) \leq \Aff(\Omega)$, we have a natural duality between $\V(A)$ and $\E(A)$, or, to put it another way, there is a natural linear mapping $\V(A) \rightarrow \E(A)^{\ast}$, taking each $\alpha \in \Omega$ to the corresponding evaluation functional in $\E(A)^{\ast}$. Since states are, for us, probability weights on $X(A)$, this mapping 
is injective. \\


\noindent{\bf State-Completeness} If $A = (X,\Omega)$ is a model, with
linear hull $\E(A)$, then any positive linear functional $\alpha \in
\E(A)^{\ast}$ with $\alpha(u) = 1$ (that is, any state {\em on} $\E$)
defines a probability weight on $X(A)$ by restriction. Let
$\widehat{\Omega}$ denote the set of such states. Obviously, $\Omega
\subseteq \widehat{\Omega}$. We may regard $\widehat{\Omega}$ as the
set of probability weights that are consistent with all of the linear
relations among outcomes that are satisfied by the given state space
$\Omega$. Evidently, the assignment $\Omega \mapsto \widehat{\Omega}$
is a closure on the poset of closed convex subsets of
$\Bomega(X)$. Call a model {\em state-complete} iff $\Omega =
\widehat{\Omega}$.

\begin{lemma} Let $A = (X,\Omega)$ be a finite-dimensional probabilistic model. Then the following are equivalent: 
\begin{itemize} 
\item[(a)] $A$ is state-complete
\item[(b)] $\E(A)_{+} = \E(A) \cap \Aff_{+}(\Omega) = \E(A) \cap \V(A)^{\ast}$; 
\item[(c)] The canonical mapping $\V(A) \rightarrow \E(A)^{\ast}$ is surjective, hence, an order-isomorphism.
\end{itemize} \end{lemma} 

\noindent \noindent{\em Proof:} To see that (a) implies (b), suppose $f \in \Aff_{+}(\Omega) \setminus \E(A)_{+}$. Then (by the finite-dimensional version 
of the Hahn-Banach separation theorem) there exists some $\alpha \in \E(A)^{\ast}$ with $\alpha(a) \geq 0$ for all 
$a \in \E(A)_+$ but $\alpha(f) < 0$. We can normalize $\alpha$ so that $\alpha(u) = 1$, in wich case $\alpha \in \widehat{\Omega}$. Since $f$ is non-negative on $\Omega$, it follows that $\alpha \not \in \Omega$, whence, 
$\widehat{\Omega} \not = \Omega$, and $A$ is not state-complete. Conversely, if $\alpha \in \widehat{\Omega} \setminus \Omega$, 
then we can find some $f \in \E(A)^{\ast \ast} = \E(A)$ with $f(\alpha) < 0$ but $f(\beta) \geq 0$ for all $\beta \in \Omega$. But 
now $f \in \E \cap \Aff_{+}(\Omega)$, and yet --- as $a(\alpha) \geq 0$ for all $a \in \E(A)_{+}$ --- we have 
$f \not \in \E(A)_{+}$. Thus, (b) implies (c).  As all systems here are finite-dimensional, (b) and (c) are clearly equivalent. $\Box$ \\



\noindent{\bf \emph{Standing Assumption:}} {\em Henceforth,} {\bf \emph{all models are state-complete}}. \\

One might almost, at this point, regard the test space $X(A)$ as
merely a sort of builder's scaffolding, to be discarded once the space
$\E(A)$ has been constructed. For many applications, this works
perfectly well. However, the additional structure represented by $X$
turns out to be useful in many ways, so we prefer to retain it for present 
purposes
Doing so imposes no additional restrictions on the structure of $\E(A)$ 
because, given an order-unit space $\E$, we can always take $X$ to
consist of {\em all} observables on $\E$, as discussed above. \footnote{One of many uses for the test space structure is to privilege certain 
classes of observables on an order-unit space having special order-theoretic 
properties --- for example, the set of observables the outcomes of which lie on 
extremal rays of $\E_+$ forms a test space, or those whose outcomes are 
atomic effects, i.e., those that lie on extremal rays of $\E_+$ {\em and} are 
extreme points of $[0,u]$. }\\



\noindent{\bf Direct Sums of Models} 
A {\em face} of a convex set $K$ is a convex subset 
$J \subseteq K$ such that, for all $a, b \in K$ and all $0 \leq t \leq 1$, 
\[ta + (1-t)b \in J \ \Rightarrow \ a \in K \ \mbox{and} \ b \in K.\]
If $J$ and $K$ are cones, then this is equivalent to the condition
that $a + b \in J \ \Rightarrow \ a \in J$ and $b \in J$. A minimal
face of a cone is in fact a ray; we more usually speak of an {\em
  extremal ray}. An element of a cone is {\em ray-extremal}, or simply
{\em extremal}, iff it generates an extremal ray. In finite
dimensions, every (closed) cone is the convex hull of its extremal
elements.

The {\em direct sum} of two ordered vector spaces $\E$ and $\E$ is 
their vector-space direct sum, $\E \oplus \F$, equipped with the cone 
$\E_+ \oplus \F_+$ consisting of all sums of positive elements from each. This is the smallest cone in $\E \oplus \F$ making the standard embeddings $\E, \F \rightarrow \E \oplus \F$ given by $a \mapsto (a,0)$ and $b \mapsto (0,b)$ (for $a \in \E$ and $b \in \F$) positive. In this case, $\E_+$ and $\F_+$ are both faces of $\E_+ \oplus \F_+$. .. $\E$ is irreducible iff not a direct sum. 

If $X$ and $Y$ are sets, we write $X \oplus Y$ for their coproduct (or disjointified union), 
\[X \oplus Y = \{1\} \times X \cup \{2\} \times Y.\]
If $X$ and $Y$ are test spaces, we make $X \oplus Y$ into a test space by letting $\A(X \oplus Y)$ equal the 
set $\{ E \oplus F | E \in \A(X), F \in \A(Y)\}$. 
We can understand a test of the form $E \oplus F$ as a two-stage test: first, perform the classical two-outcome test $\{1,2\}$ (by flipping a coin, say); if the result is $1$, measure $E$, if the result is $2$, measure $F$. A probability weight $\omega$ on $X \times Y$ corresponds to an arbitrary choice of a probability weight $p$ on $\{1,2\}$ and probability weights 
$\alpha \in \Omega(X)$ and $\beta\in \Omega(Y)$, by 
\[ \omega(1,x) = p(1) \alpha(x) \ \ \text{and} \ \ \omega(2,y) = p(y)\beta(y).\]
The weights $p$, $\alpha$ and $\beta$ are uniquely determined by $\omega$, so we can unambiguously write 
\[\omega = t \alpha + (1 - t)\beta\]
In other words, $\Omega(X \oplus Y) = \Omega(X) \oplus \Omega(Y)$, whence, $\E(X \oplus Y) = \E(X) \oplus \E(Y)$. 

Every discrete classical probablistic model $(E,\Delta(E))$ is a direct convex sum of trivial models $(\{x\},\delta_{x})$ where $x \in E$ and $\delta_{x}(x) = 1$. 
In contrast, the basic quantum model $(X(\H),\Omega(\H))$ is irreducible. The more general models associated 
with matrix algebras arise as direct sums of irreducible quantum models. \\ 

\subsection{Processes and Categories}  

In very broad terms, a {\em probabilistic theory} might be nothing
more than a class of probabilistic models. But this usage is really
much too broad. Part of the job of a theory is to tell us, not only
which models represent ``actual" systems, but also something about how
such systems can change. In order to speak about systems changing, we
need to introduce into the preceding formalism a notion of {\em
  process}. A natural place to start is with the idea of a mapping
$\phi : \alpha \mapsto \phi(\alpha)$ taking states $\alpha$ of an
initial (or input) system $A$ to states of a final (output) system
$B$. To allow for ``lossy" processes or conditioning, we should permit
$\phi(\alpha)$ be be a sub-normalized state of $B$ when $\alpha$ is a
normalized state of $A$. Finally, since randomizing the input state
should randomize the output state in the same way, we should expect
this $\phi$ be an affine mapping. Thus, we model a process from $A$ to
$B$ by an affine mapping $\phi : \Omega(A) \rightarrow \E(B)$ with
$u_{B}(\phi(\alpha)) \leq 1$; or, what is the same thing, by a
positive linear mapping $\phi : \E(A)^{\ast} \rightarrow \E(B)^{\ast}$
with $u_{B} \circ \phi \leq u_{A}$. We can interpret
$u_{B}(\phi(\alpha))$ as the {\em probability} that $\phi$ occurs when
the initial state is $\alpha$ --- or, perhaps more accurately, as the 
probability that the process occurs, {\em if} initiated.


To every process $\phi : \V(B) \rightarrow \V(A)$, there corresponds a
{\em dual process} $\tau = \phi^{\ast} : \E(A) \rightarrow \E(B)$,
given by $\phi^{\ast}(a) = a \circ \phi$ for any $a \in
\E(A)$. Operationally, to measure $\phi^{\ast}(a)$ on a state
$\alpha$, one first subjects the state $\alpha$ to the process $\phi$,
and then makes a measurement of the effect $a$. Note that
$\tau(u)(\alpha) = u(\tau^{\ast}(\alpha))$ is the probability that the
process $\tau^{\ast} = \phi$ occurs if the initial state is
$\alpha$. In what follows, it will often be more convenient
mathematically to deal with these dual processes. In other words, to
use physicists' lingo, we'll often work with the ``Heisenberg" rather
than the ``Schr\"odinger" picture of processes.

Not every positive linear mapping $\V(A) \rightarrow \V(B)$ will
generally count as a process. As remarked above, it is part of the job
of a probabilistic theory to specify those that do. However, it seems
reasonable to require that convex combinations of processes and
composites of (composable) processes also count as processes. It will
also be convenient to assume that, for every pair of systems $A$ and
$B$, there is a {\em null process} that takes every state $\alpha \in
\Omega(A)$ to the {\em zero state} $0 \in \E(B)$. It seems reasonable,
also, that there exist a canonical {\em trivial} sytem $I$,
corresponding to a test space with only a single outcome, $1$, and a
single test $\{1\}$. We then have $\E(I) = \E(I)^{\ast} = \R$. We can
then require that, for every normalized state $\alpha \in \V(A)$,
there exist a process $\R \rightarrow \V(A)$ of {\em preparation},
given by $1 \mapsto \alpha$, and, for every outcome $x \in X(A)$, a
process $\V(A) \rightarrow \R$ of {\em registration}, sending $\alpha
\in \V(A)$ to $\alpha(x)$. The dual process corresponding to the
preparation of $\alpha$ is simply the state $\alpha$ itself, while the
process dual to the registration of $x$ is the linear mapping $\R
\rightarrow \E(A)$ sending $1$ to $x$. All of this suggests the
following


\begin{definition} {\em A (state-complete) {\bf probabilistic theory}\footnote{This definition differs from that of \cite{BW09}, most obviously in that objects are associated with effect spaces, rather than state spaces, but also in taking the test 
space $X(A)$ to be part of the structure of $A \in \C$.} is a category $\cc$ such that 
\begin{itemize} 
\item[(1)] Every object $A \in \C$ is a probabilistic model;
\item[(2)] For all $A, B \in \C$, the set $\C(A,B)$ of morphisms $A \rightarrow B$ is a closed, convex subset of $\L_{+}(\E(A),\E(B))$, containing the zero mapping, and 
with $\tau(u_A) \leq u_B$ for all $\tau \in \C(A,B)$; 
\item[(3)] There is a distinguished {\em trivial system} $I$ with $\E(I) = \R$ and $X = \{1\}$, such that 
for every $A \in \C$, $X(A) \subseteq \C(I,A)$ and $\Omega(A) \subseteq \C(A,I)$.
\item[(4)] The order unit $u_A \in \E(A)$ belongs to $\C(I,A)$. 
\end{itemize}}
\end{definition} 




\noindent{\bf \emph{From now on, we work in a fixed probabilistic
    theory $\C$ of this kind.} }  We write $\C^{\ast}$ for the
category having the same objects, but with morphisms $\C^{\ast}(A,B)$
the set of mappings $\phi = \tau^{\ast} : \V(B) \rightarrow \V(B)$
with $\tau \in \C(B,A)$.  In effect, $\C$ and $\C^{\ast}$ offer,
respectively, the ``Heisenberg" and the ``Schr\"odinger" picture of the
same theory. Depending on context, we shall understand the word
``process" to refer either to a morphism $\tau \in \C(A,B)$ for some
$A, B \in \C$, or to the dual mapping $\phi = \tau^{\ast} : \V(B)
\rightarrow \V(A)$.

\begin{example} 
{\em By a {\em standard finite-dimensional quantum theory}, we mean a
  category $\C$ of probabilistic models $(\E,X)$ where $\E$ is the
  hermitian part of a finite-dimensional complex matrix algebra (a
  direct sum of algebras of the form $\L(\H)$), with
  trace-nonincreasing completely positive mappings as morphisms. In
  this formulation, classical probabilistic theories arise as the
  degenerate case in which all of the matrix algebras associated with
  systems in $\C$ are commutative.} \end{example}





\noindent{\bf Reversible and Probabilistically Reversible Processes} A process
$\tau \in \C(A,B)$ is {\em reversible} iff it is invertible as a
morphism in $\C$, i.e., there exists an inverse process $\tau^{-1} \in
\C(B,A)$ with $\tau^{-1} \circ \tau = \id_{A}$ and $\tau \circ
\tau^{-1} = \id_{B}$. In this case, $\tau$ is an order-automorphism
$\E(A) \simeq \E(B)$, and $\tau^{-1} : \E(B) \simeq \E(A)$ is the
inverse isomorphism. Moreover, for such a process, we have
$\tau(u_{A}) = u_{B}$: by assumption, $\tau(u_{A}) \leq u_{A}$, and
also $\tau^{-1}(u_{B}) \leq u_{A}$, whence, as $\tau$ preserves order,
$u_{B} \leq \tau(u_{A})$.  Dually, a process $\phi \in \C^{\ast}(A,B)$
is reversible iff it has an inverse in $\C^{\ast}(B,A)$; equivalently,
$\phi$ is invertible iff the dual process $\tau = \phi^{\ast}$ is
invertible. In this case, we have $u_{B}\phi(\alpha) = 1$ for every
normalized state $\alpha \in \Omega(A)$.

There is a weaker but very useful notion, which we shall call {\em
  probabilistic reversibility}. This is slightly easier to describe in
terms of processes acting on states, rather than effects:

\begin{definition} {\em A process $\phi \in \C^{\ast}(A,B)$, 
is {\em probabilistically reversible} iff it is invertible as a linear
mapping $\V(A) \rightarrow \V(B)$, with a positive inverse {\em and}
if the inverse mapping $\phi^{-1}$ is a positive multiple of a process
$\phi_{o} \in \C^{\ast}(B,A)$ --- say, $\phi^{-1} = c \phi_o$ with $c
> 0$. }\end{definition}
Operationally, this means that there is some non-zero probability that $\phi_o \circ \phi$ will return the system
to its original state.  Indeed,
\[\phi_o(\phi(\alpha))(u_{A}) = c^{-1}\phi^{-1}(\phi(\alpha))(u_{A}) = c^{-1} \alpha(u_{A}) = c^{-1},\]
so this probability is exactly $1/c$.  In particular, $\phi$ is
reversible with probability one iff $c = 1$, so that $\phi^{-1}$ is a
process in $\C^{\ast}(B,A)$ --- in other words, $\phi$ is an reversible process. 

We shall say that a process $\tau \in \C(A,B)$ is reversible with
probability $1/c$ iff $\tau^{\ast} \in \C^{\ast}(A,B)$ is
reversible. Obviously, the set of probabilistically reversible
processes, in either $\C(A,A)$ or $\C^{\ast}(A,A)$, is a group,
containing, but larger than, the group of all reversible processes on
$A$.\\

\noindent{\bf Historical remarks:} The representation of what we are calling probabilistic models in terms of an order-unit space and its dual goes back at least to the work of Davies and Lewis \cite{Davies-Lewis} and Edwards \cite{Edwards}. A 
good survey of the relevant functional analysis can be found in \cite{Alfsen-Shultz}. 
Test spaces --- originally called ``manuals" --- were the basis for a generalized probability theory (and an associated  ``empirical logic") developed in the 1970s and 80s by C. H. Randall and D. J. Foulis and their students. See \cite{Wilce10} for a survey. Mathematically, of course, a test space is just a hypergraph; the current terminology serves only to reinforce the intended probabilistic interpretation.  \\

\section{Composition and Entanglement} 


Consider two systems, $A$ and $B$, which are not interacting in any
obvious, causal sense -- for example, systems occupying space-like
separated regions of space-time. In this situation, it seems
reasonable to assume that what that can be {\em happen} to each system
idividually --- the preparation of a state, the making of a
measurement, etc. --- can happen together, independently.

Another natural (albeit more contingent) requirement is a {\em
  no-signaling} condition, forbidding the transmission of information
from $A$ to $B$, or vice versa, by the mere decision to make one
measurement rather than another on $A$, or on $B$.  As we'll see, the
phenomenon of {\em entanglement}, one of the supposed hallmarks of
quantum theory, is actually a rather generic feature of such
``non-signaling" composite systems in non-classical probabilistic
theories, whether "quantum" or otherwise. (Indeed, the phenomenon even
arises in otherwise quite classical theories involving a restricted
set of probability weights.)

\subsection{Composites of Models} 

Suppose two parties --- Alice and Bob, say --- control, respectively, systems $A$ and $B$, which occur as components of some composite system $AB$, but are still sufficiently isolated to be prepared and measured separately.  
At a very minimum, we would expect Alice's making a measurement, $E$, on here part of the composite system, and Bob's making a measurement, $F$, on his part, {\em constitutes} the making of a measurement on the combined system. 
We would also expect that states of the two component systems can be prepared independently. Formalizing these requirements, 
we arrive at the following: 



\begin{definition}\label{def: composite}  A {\em composite} of two probabilistic models $A$ and $B$ is a model 
$AB$, together with a mapping 
\[X(A) \times Y(B) \rightarrow X(AB): \ \ (x,y) \mapsto xy\] 
such that 
\begin{itemize}
\item[(i)] for all tests $E \in \A$ and $F \in B$, the {\em product test} 
$EF := \{ xy | x \in E, y \in F \}$ 
belongs to $\A(AB)$; and 
\item[(ii)] for all states $\alpha \in \Omega(A)$ and $\beta \in \Omega(B)$, there exists a {\bf unique [?]} state $\alpha \otimes \beta \in \Omega(AB)$ with 
$(\alpha \otimes \beta)(xy) = \alpha(x)\beta(y)$.
\end{itemize} 
\end{definition}

\noindent{\em Remarks:} There are several ways in which we might
plausibly weaken this definition. For instance, we might require only
that the product outcome $xy$ be an {\em effect} in $\E(AB)_+$, and the set
$EF$, an observable, but not necessarily a test,  of 
$AB$.  \footnote{More radically, one might consider models of systems
  interacting in such a way that the making of a particular
  measurement, or the preparation of a particular state, on one
  component, {\em precludes} the making of certain measurements, or
  the preparation of certain states, on the other
  component. Mathematically, such situations are certainly possible.}
Such possibilities are worth bearing in mind. However, for the
purposes of this survey, it seems reasonable to use the more
restrictive, but therefore simpler, definition above.  Note in (ii) we
require only the existence, but not the uniqueness {\bf [??]}, of product states
(where a product state for $\alpha$ and $\beta$ is defined as a state
$\gamma$ with $\gamma(xy) = \alpha(x)\beta(y)$, and a product state
\emph{tout court} as one that is a product state for a pair of states
$\alpha$ and $\beta$).  \\


The injectivity of the mapping $x,y \mapsto xy$ in condition (i) allows us to identify $X(A) \times X(B)$ with the Let us write 
\[X(A)X(B)\ := \ \{ xy | x \in X, y \in Y\}\] 
for the square of {\em product outcomes} in $Z$. With a slight abuse of notation, we may write $\A(A) \times \A(B)$ for the test space consisting of product tests $EF$. Condition (i) asserts that $\A(A) \times \A(B)$ is contained in $\A(AB)$, so every state in $\Omega(AB)$ restricts to a state $\omega_o$ on the former.  
Where the restricted state $\omega_o$ {\em determines} the global
state $\omega$ --- that is, where the set $X(A)X(B)$ of product
outcomes is state-separating --- we say that the composite is {\em
  locally tomographic}. In this setting, the joint probabilities of
outcomes of measurements on the component systems $A$ and $B$,
completely determine the state of the composite.\footnote{Barrett
  \cite{Barrett} calls this the {\em global state hypothesis}; the
  term {\em locally tomographic} seems to have become more standard.}
This is a reasonable, but also a rather strong, restriction. Indeed,
while composites in standard complex QM are locally tomographic, this
is not the case for real or quaternionic QM. We'll return to this
matter below.

\begin{example}[Composite quantum models]{\em If $A(\H)$ and $A(\K)$ are two quantum-mechanical models, associated with finite-dimensional Hilbert spaces $\H$ and $\K$, respectively, let 
\[A(\H)A(\K) \ = \ A(\H \otimes \K)\]
the model associated with $\H \otimes \K$. That is, 
$\A(\H \otimes \K)$ consists of orthonormal bases for $\H \otimes \K$, 
while 
$\Omega(\H \otimes \K)$ consists of density operators on $\H \otimes \K$. If $x \in \H$ and $y \in \K$ are unit vectors, 
then $x \otimes y$ is a unit vector in $\H \otimes \K$. It is easy to check that $x, y \mapsto x \otimes y$ 
makes $A(\H \otimes \K)$ into a composite in the sense of the preceding definition. 
 }\end{example}

\subsection{Non-Signaling Composites and Entanglement}  The very broad definition of a composite system given above leaves room for situations in which the probability of Bob's obtaining 
an outcome $y$ will depend on which test $E \in \A(A)$ Alice chooses
to measure. This is plausible only in scenarios in which Alice's
measurements are able physically to disturb Bob's system. If we wish
to model composites in which the two systems $A$ and $B$ are
suffciently isolated from one another that this kind of remote
disturbance is ruled out ---- the obvious situation being one in which
$A$ and $B$ are spacelike separated --- then we must impose a further
constraint.

\begin{definition} A probability weight $\omega$ on $\A(A) \times \A(B)$ is {\em non-signaling} iff it has well-defined {\em marginal} (or {\em reduced}) states, in the sense that 
\[\omega_1(x) := \sum_{y \in F} \omega(xy) \ \ \text{and} \ \  \omega_2(y) := \sum_{x \in E} \omega(xy)\]
are independent of the choice of tests $E \in \A(A)$, $F \in \A(B)$. \end{definition}

If $\omega \in \Omega(AB)$ is non-signaling, then for every $y \in X(B)$ and $x \in X(A)$, we can define the {\em conditional states} $\omega_{1|y}$ and $\omega_{2|x}$ on $A$ and $B$, respectively, by 
\[\omega_{1|y}(x) := \frac{\omega(xy)}{\omega_2 (y)} \ \text{and} \ \omega_{2|x}(y) := \frac{\omega(xy)}{\omega_1(x)}.\]
These are well-defined probability weights on $\A(A)$ and $\A(B)$,
respectively. It would seem reasonable to include them in the state
spaces of $A$ and $B$.
Therefore, we adopt the following language:

\begin{definition}{\em A {\em non-signaling composite} of $A$ and $B$ is a composite $AB$ in which all 
states are non-signaling, {\em and} all conditional states belong to
the designated state spaces of $A$ and $B$ --- that is, $\omega_{2|x}
\in \Omega(B) \ \ \text{and} \ \ \omega_{1|y} \in \Omega(A)$ for all
$x \in X(A)$ and $y \in X(B)$. }\end{definition}

This has a strong consequence \cite{Wilce92}:

\begin{lemma}[Bi-Linearization] 
Let $AB$ be a non-signaling composite of $A$ and $B$.  Then every state 
$\omega \in \Omega(AB)$ extends uniquely to a bilinear form on $\E(A) \times \E(B)$. \end{lemma} 

\noindent{\em Proof:} For every $x \in X(A)$, define  $\hat{\omega}(x) \in \R^{X(B)}$ by $\hat{\omega}(x)(y) = \omega(x,y)$. 
Notice that $\omega_{2|x} = \hat{\omega}(x)/\omega_1(x)$. Since the conditional state $\omega_{2|x}$ belongs 
to $\Omega(B)$, we have $\hat{\omega}(x) \in \V(B) = \E(B)^{\ast}$, with $\sum_{x \in E} \hat{\omega}(x) = \omega_{2}$. Dualizing (and remembering that $\E(A)$ is finite-dimensional), we have a linear mapping $\hat{\omega}^{\ast} : \E(B) \rightarrow \R^{X(A)}$. Now, $\hat{\omega}^{\ast}(y) = \omega_{1|y}/\omega_{2}(y)$; the latter belongs to $\Omega(A)$, so $\hat{\omega}^{\ast}(y) \in \V(A) = \E(A)^{\ast}$ for every $y \in X(B)$. Since $X(B)$ spans $\E(B)$, it follows that the range of $\hat{\omega}^{\ast}$ lies in $\V(A)$, i.e., we can regard $\hat{\omega}^{\ast}$ as 
a linear mapping $\E(A) \rightarrow \V(B) = \E(B)^{\ast}$. Equivalently, we have a bilinear form ${\mathcal B}_{\omega}(a,b) = \hat{\omega}^{\ast}(b)(a)$, which evidently satisfies $B_{\omega}(x,y) = \omega(xy)$ for all 
$x \in X(A), y \in X(B)$. Since $X(A)$ and $X(B)$ span $\E(A)$ and $\E(B)$, the form ${\cal B}_{\omega}$ is uniquely 
determined by this property. $\Box$\\ 

It follows that, for a non-signaling composite, the mapping $X(A) \times X(B) \rightarrow X(AB): x,y \mapsto xy$ 
gives rise to a linear mapping  $\otimes : \E(A) \otimes \E(B) \rightarrow \E(AB)$, with $\omega(x \otimes y) = {\mathcal B}_{\omega}(x,y) = \omega(xy)$ for every $\omega \in \E(AB)^{\ast}$. The composite $AB$ is locally tomographic iff this mapping is surjective. 


\begin{corollary} A non-signaling composite $AB$ of models $A$ and $B$ is locally tomographic iff $\E(AB) \simeq \E(A) \otimes \E(B)$, that is, $\dim(\E(AB)) = \dim(\E(A))\dim(\E(B))$.
\end{corollary}

Lemma 3 allow us to extend the definition of conditional states to arbitrary effects, setting 
\[\omega_{1|b}(a) = \omega(a \otimes b)/\omega(u \otimes b) \ \ \mbox{and} \ \ \omega_{2|a}(b) = \omega(a \otimes b)/\omega(a \otimes u)\]
for arbitrary effects $a \in \E(A)$ and $b \in \E(B)$ 
(with the usual proviso about division by zero). The following bipartite version of the law of total probability is easily verified:

\begin{lemma}[Law of Total Probability] Let $AB$ be a non-signaling composite of $A$ and $B$; let $\omega$ be 
any state on $AB$, and let $E$ and $F$ be any two observables on $\E(A)$ and $\E(B)$, respectively, 
then 
\[\omega_{2} = \sum_{a \in E} \omega_{1}(a) \omega_{2|a} \ \ \mbox{and} \ \ \omega_1 = \sum_{b \in F} \omega_{2}(b) \omega_{1|b}\]
\end{lemma}

\begin{corollary} Let $AB$ be a non-signaling composite of $A$ and $B$, and let $\omega$ be a pure state of $AB$. If
the marginal state $\omega_2$ is pure, then $\omega_1$ is also pure,
and $\omega = \omega_1 \otimes \omega_2$. \end{corollary}

\noindent{\em Proof:} It is easy to see that, if a product state $\omega = \omega_1 \otimes \omega_2$ is pure, then 
both marginals must be pure. Now suppose that one marginal state --- say, $\omega_2$ --- is pure. Since $\omega_2 = \sum_{x \in E} \omega_1(x)\omega_{2|x}$, and the conditional states $\omega_{2|x}$ belong to $\V(B)$, it follows that for every $x \in E$ with $\omega_1(x) > 0$, we must have
$\omega_{2|x} = \omega_2$, so that $\omega(xy) = \omega_{1}(x)\omega_2(y)$ for every such $x$. The same result holds trivially if $\omega_1(x) = 0$, so we have $\omega(xy) = \omega_1(x) \omega_2(y)$ for all choices of $x$ and $y$. It follows that 
$\omega = \omega_1 \otimes \omega_2$. $\Box$\\ 

\begin{definition}{\em A state $\omega$ on $AB$ is {\em separable} iff it is a mixture of product states, that is, $\omega = \sum_i t_i \alpha_i \otimes \beta_i$ where $t_i \geq 0$ and $\sum_i t_i = 1$
A state {\em not} of this form is said to be {\em entangled}.} \end{definition} 

Using this language, the preceding Corollary gives us 

\begin{corollary} If $AB$ is a non-signaling composite of models $A$ and $B$, and 
$\omega$ is an entangled state of $AB$, then both $\omega_1$ and
  $\omega_2$ are mixed. \end{corollary}

This is often regarded as the hallmark of entangled {\em quantum}
states; but, as we see, it is really a quite general possibilty
arising in any non-classical probabilistic setting.  Of course, one
can still ask at this point whether entangled states {\em exist} in
any generality, once one leaves the confines of quantum
theory. However, as we'll see in Section 3.4 below, there is a sense
in which {\em most} non-signaling composites of non-classical models
admit entangled states. \\

\noindent{\bf The CHSH Inequality} 
Let $AA$ be a non-signaling composite of two copies of $A$. For any $a, b \in \E(A)$ with $-u_A \leq a, b \leq u_A$, let 
$a' = u_A - a$ and $b' = u_A - b$. For any state $\omega$ in $AA$, define 
\[S(\omega; a,b) = \omega(a,b) + \omega(a,b') + \omega(a',b) - \omega(a',b').\]
This is called the CHSH (Clauser-Horn-Shimony-Holt) parameter associated with $\omega$, $a$ and $b$. of a bipartite
If $\omega$ is a product state, then $S \leq 2$ for all choices of $a$ and $b$; as $S$ is affine in $\omega$, it follows 
that $S \leq 2$ for all separable states. For entangled states it can be larger. A priori, the upper bound for $S$ is $4$, and 
this is achieved, for example, if $A$ is the ``square bit" of example 3. However, for
bipartite quantum states, the upper bound is much lower. As pointed out by Tsirel'son \cite{Tsirel'son}, 
$S \leq 2\sqrt{2}$ for any quantum bipartite state and any effects $a$ and $b$. A great deal of work has gone into trying to find a deeper explanation for this bound. \cite{Allcock, InfCausality}. In section 4, we will return to this matter. \\


\noindent{\bf Conditioning Maps and Isomorphism States} If $\omega$ is any non-signaling state on $AB$, then the associated bilinear form $\Bi_{\omega}$ on $\E(A) \times \E(B)$ gives us a positive linear mapping 
\[\widehat{\omega} : \E(A) \rightarrow \E(B)^{\ast}\]
defined by 
\[\widehat{\omega}(a)(b) = \omega(a \otimes b)\]
for all $a \in \E(A)$ and $b \in \E(B)$. 
Notice that $\widehat{\omega}(a) = \omega_{1}(a) \omega_{2|a}$. Accordingly, we think of $\widehat{\omega}(a)$ as an {\em un-normalized conditional state} of $B$ given the effect $a \in \E(A)$, and refer to $\widehat{\omega}$ as the {\em conditioning map} associated with $\omega$. Of course, there is also a conditioning map running in the opposite direction. In fact, this is just the adjoint of $\widehat{\omega}$; that is, 
 $\widehat{\omega}^{\ast}(b)(a) = \widehat{\omega}(a)(b) = \omega(a,b)$ for all effects $a \in \E(A)$ and $b \in \E(B)$. 

There is a dual construction for effects. An effect $f \in \E(AB)$ defines a positive bilinear form on $\V(A) \times \V(B)$ by $(\alpha,\beta) \mapsto f(\alpha \otimes \beta)$. This, in turn, yields a positive linear mapping 
\[\hat{f} : \V(A) \rightarrow \V(B)^{\ast} = \E(B)\] 
given by $\hat{f}(\alpha)(\beta) = f(\alpha \otimes \beta)$. We call 
$\hat{f}$ the {\em co-conditioning map} associated with $f$.

\begin{definition}{\em Let $AB$ be a non-signaling composite of $A$ and $B$. An {\em isomorphism state} on $AB$ is a state $\omega \in \Omega(AB)$ such that the conditioning map $\hat{\omega} : \E(A) \rightarrow \V(B)$ is an order-isomorphism. Dually, an {\em isomorphism effect} is an effect $f \in \E(AB)$ such that the co-conditioning map $\hat{f} : \V(A) \rightarrow \E(B)$ is an order-isomorphism.}\end{definition} 

Evidently, the inverse of an isomorphism state is a multiple of an
isomorphism effect, and vice versa. This point will be important in
the discussion of teleportation protocols below.  If there exists an
isomorphism state on a composite $AA$ of $A$ with itself, then we have 
$\E(A) \simeq \V(A) = \E(A)^{\ast}$.\footnote{The converse is not quite true:
  an order-isomorphism $\E(A) \simeq \V(A)$ defines a non-signaling
  state on $A \maxtensor B$ {\bf [def.]}, but need not correspond to a state of
  $AB$.} 
More generally, we shall say that $A$ is  {\em weakly self-dual} iff there exists 
an order-isomorhism $\E(A) \simeq \V(A)$ (equivalently: an isomorphism state in $A \maxtensor A$). 
Although this is a strong constraint on the structure
of a probabilistic model, it is nevertheless satisfied by many
examples that are neither quantum nor classical. For example, the
models associated with state spaces that are regular $2$-dimensional
polytopes --- that is, regular $n$-gons --- are weakly self-dual.  

As we'll discuss further in Section 5, quantum models satisfy a much stronger form of
self-duality: not only does there exist an order-isomorphism $\V(\H) \simeq \E(\H)$, but 
this is given by an inner product on $\E(\H) = \L(\H)$, namely, $a \mapsto \Tr(a \cdot)$. 

\begin{proposition}[\cite{BGW09}]\label{prop: isomorphism states pure} 
Let $A$ and $B$ be irreducible, and let $AB$ be any
locally-tomographic, non-signaling composite of $A$ with $B$. Then any
isomorphism state in $AB$ is pure in $\Omega(AB)$, and any isomorphism
effect is extremal in $\E(AB)_{+}$. \end{proposition}

If  $A$ and $B$ are not irreducible, an isomorphism state on $AB$ need not be pure. 
For example, if $A = B = (E,\Delta(E))$, then any state uniformly correlating $A$ and $B$ --- 
say $\omega(x,x) = 1/|E|$ and $\omega(x,y) = 0$ for $x \not = y$ --- is an isomorphism state, 
but will be pure only if $|E|  = 1$. 



\subsection{Quantum Composites} 

This is a good place at which to pause for a second and more detailed look at quantum-mechanical composites. 
As noted earlier in Example \ref{ex: quantum models}, the mapping $X(\H) \times X(\K) \mapsto X(\H \otimes \K)$ given by $x, y \mapsto x \otimes y$ turns $A(\H \otimes \K)$ into a composite of the models $A(\H)$ and $A(\K)$. This mapping extends to the  bilinear mapping  
\[\E(\H) \times \E(\K) = \L_{h}(\H) \times \L_{h}(\K) \rightarrow \L_{h}(\H \otimes \K) = \E(\H \otimes \K),\]
that sends $a, b \in \L_h(\H) \times \L_h(\K)$ to the operator $a \otimes b$ on $\H \otimes \K$ (given by $(a \otimes b)(x \otimes y) = ax \otimes by$ for all $x \in \H$, $y \in \K$). Hence, by  
Lemma 3, $A(\H \otimes \K)$ is a {\em non-signaling} product of $A(\H)$ and $A(\K)$. \\

\noindent{\bf Conditioning} 
Let $\H$ be a complex Hilbert space. For any vectors $x, y \in \H$, let $x \odot y$ denote  the  rank-one operator on $\H$ 
given by $(x \odot y) z = \langle z, y \rangle x$. (In Dirac notation, this is $|x \rangle \langle y |$.) 
If $x$ is a unit vector, then $x \odot x = P_x$, the orthogonal projection operator associated with $x$. 

The mapping $x, y \mapsto x \odot y$ is sesquilinear, that is, linear in its first, and conjugate linear in its second, argument; it therefore extends to a linear mapping 
$\H \otimes \bar{\H} \rightarrow \L(\H)$, 
where $\bar{\H}$ is the conjugate space of $\H$, 
taking any vector $v = \sum_{i} t_{i} x_i \otimes \bar{y}_i$ to the corresponding operator $\hat{v} := \sum_{i} t_i x_i \odot y_i$. It is easy to see that this is injective and hence, on dimensional grounds, an isomorphism. It is 
useful to note that 
\[\langle \hat{v}(x),y \rangle = \langle v, y \otimes \bar{x}\rangle\]
for all $x, y \in \H$. Hence, if $v$ is any unit vector in $\H \otimes \bar{\H}$, the corresponding pure state $\omega = \alpha_{v}$ of $A(\H \otimes \bar{\H})$ assigns joint probabilities to outcomes $x \in X(\H)$ and $\bar{y} \in X(\bar{\H})$ by 
\[\omega(x,\bar{y}) = \left | \langle v, x \otimes \bar{y} \rangle \right |^2 = \left | \langle \hat{v}(y), x \rangle \right |^2\]
so that the conditional state $\omega_{2|\bar{y}}$ is exactly the pure state associated with the unit vector 
$\hat{v(y)}/\|\hat{v(y)}\|$. (The fact that conditioning a pure bipartite quantum state by a measurement outcome always leads to a pure state --- the {\em pure conditioning property} --- is rather special, and has 
been exploited in \cite{CDP, Wilce12}.) \\


\noindent{\bf Purification and Correlation} Suppose now that $\alpha$ is a state 
on $A(\H)$, represented by a density operator $W$  on $\H$ with spectral  
resolution 
\[W = \sum_{x \in E} \lambda_x P_x = \sum_{x \in E} \lambda_{x} x \odot x\]
where $E$ is an orthonormal basis for $\H$ and $\sum_{x \in E} \lambda_{x} = \Tr(W) = 1$. Functional calculus gives us $W^{1/2} = \sum_{x \in E} \lambda^{1/2}_{x} x \odot x$. 
We can interpret this as a unit vector in $\H \otimes \bar{\H}$, namely 
\begin{equation}\Psi_{W} := \sum_{x \in E} \lambda^{1/2}_{x} x \otimes \bar{x}.\end{equation}
This, in turn, defines a bipartite state on the composite quantum system $A \bar{A} := A(\H \otimes \overline{\H})$. The marginal, or reduced, state of the first component system is 
given by 
\[\omega_1(a) = \Tr(P_{\Psi_{W}} (a \otimes \1_{\overline{\H}}) = \langle (a \otimes \1_{\bar{\H}}) \Psi_{W}, \Psi_{W} \rangle = \Tr(Wa)\]
so the pure state corresponding to $\Psi_{W}$ is a {\em dilation} of the given mixed state $W$. 
Now observe that if $u, v \in X(\H)$ with $u \perp v$, then we have
\[\langle \Psi_{W}, u \otimes \bar{v} \rangle \ = \ \sum_{x \in E} \lambda^{1/2}_{x} \langle x, u \rangle \langle \bar{x}, v \rangle = 0.\]
Evidently, the pure state $\omega$ corresponding to $\Psi_{W}$ sets up
a {\em perfect correlation} between $E \in \M(\H)$ and the
corresponding test $\bar{E} = \{\bar{x} | x \in \E\} \in
\M(\bar{\H})$, with
\[\omega(x,\bar{x}) = |\langle \Psi_{W}, x \otimes \bar{x} \rangle|^2 = |\lambda_{x}^{1/2}|^2 = \lambda_{x}.\]

An especially interesting case arises when $\alpha$ is the maximally
mixed state, i.e., when $W = \1/n$ (where $n = \dim(\H))$. Then
$\Psi_{W}$ is {\em independent} of the choice of $E$ (since every
orthonormal basis of $\H$ is an eigenbasis for $\1$). Hence,
$\Psi_{W}$ simultaneously correlates {\em every} test $E \in \A(\H)$
with its counterpart in $\A(\overline{\H})$. Moreover, the correlation
is {\em uniform}, in that the probabilities of correlated pairs $x
\otimes \overline{x}$ of outcomes is uniformly $1/n$. As we'll see
later, the existence of such a uniformly correlating state between two
isomorphic systems has interesting consequences.\\

\noindent{\bf Local Tomography} If $\H$ and $\K$ are real or complex Hilbert spaces of dimensions $m$ and $n$, respectively, As was remarked above, $A(\H \otimes \K)$ is a non-signaling composite of $A(\H)$ and $A(\K)$. 
It is easily checked that $\dim \E(A) = \dim \Lh(\H) = m^2$ if $\H$ is
complex and $(m^2 + m)/2$ if $\H$ is real. Hence, the dimension of the
real vector space $\E(\H \otimes \K) = \Lh(\H \otimes \K)$ of
Hermitian operators is $(mn)^2 = m^2 n^2$, so in fact $\Lh(\H \otimes
\K) = \Lh(\H) \otimes \Lh(\K)$, and the composite system is locally
tomographic. On the other hand, if $\H$ and $\K$ are real, the
dimension of $\Lh(\H \otimes \K)$ is $((mn)^2 - mn)/2 + mn = ((mn)^2 +
mn)/2$, while the product of the dimensions of $\Lh(\H)$ and $\Lh(\K)$
is
\[\frac{(m^2 + m)}{2} \cdot \frac{(n^2 + n)}{2} = \frac{m^2n^2 + m^2n + mn^2 + mn}{4}.\]
This is strictly less than $(m^2n^2 + mn)/2$, which in turn is less then $(mn)^2$, so in this case, $\E(AB)$ is strictly larger than $\E(A)\otimes \E(B)$. Thus, for real Hilbert spaces $\H$ and $\K$, 
the standard composite $\A(\H \otimes \K)$ is {\em not} locally tomographic. (Neither do we have local 
tomography for quaternionic Hilbert spaces, though here, one needs to be more careful about the formulation 
of the relevant tensor products. See \cite{Baez} and \cite{KRF} for more details.)

\subsection{Maximal and Minimal Tensor Products}

Let $AB$ be a non-signaling composite of two systems $A$ and $B$.  As noted above, if $AB$ is locally tomographic, then $\E(AB) \simeq \E(A) \otimes \E(B)$ as vector spaces. In this section, we consider more closely the possibilities for such a composite. 

As we saw earlier, any non-signaling state $\omega$ on any composite
system $AB$ is associated with a bilinear form on $\E(A) \times
\E(B)$. If $AB$ is locally tomographic, then we can identify $\omega$
with this form.
We then see that there are two extreme possibilities for the set of states on a locally tomographic composite $AB$:  maximally, we may include {\em all} positive, normalized bilinear forms on $\E(A) \times \E(B)$; minimally, we may restrict our attention to the closed convex hull of the product states. 


\begin{definition} {\em Let $\E$ and $\F$ be any two finite-dimensional ordered vector spaces. The {\em minimal tensor cone} 
on $\E \otimes \F$ is the cone generated by pure tensors $a \otimes b$ with $a \in \E_+$ and $b \in \F_+$. The {\em maximal  
tensor cone} is the cone of all tensors $\tau \in \E \otimes \F$ such that $\tau(\omega) \geq 0$ for all $\omega \in \B_{+}(\E,\F)$. These two cones give us two different ordered tensor products, which we denote by 
$\E \mintensor \F$ and $\E \maxtensor \F$, respectively.} \end{definition} 

It is not difficult to see that (in finite dimensions) we have 
\[(\E \mintensor \F)^{\ast} = \E^{\ast} \maxtensor \F^{\ast} \ \mbox{and} \ (\E \maxtensor \F)^{\ast} = \E^{\ast} \mintensor \F^{\ast}.\]


Let $AB$ be any locally tomographic composite of models $A$ and $B$. 
Then the set $X(A)X(B)$ of product outcomes in $\E(AB) \simeq \E(A) \otimes \E(B)$ generates exactly 
the minimal tensor cone in $\E(A) \otimes \E(B)$. It follows that the cone of un-normalized non-signaling states on $\A(A) \times \A(B)$ [defined?] is 
exactly the maximal tensor cone in $\V(A) \otimes \V(B)$.  Dually, the set of product states generates the minimal tensor cone in $\V(A) \otimes \V(B)$. 

\begin{definition} {\em Thus, we may define the {\em minimal tensor product} of $A$ and $B$ to be the model $A \mintensor B = (\E(A) \mintensor \E(B), X(A) \times X(B))$. By the {\em maximal 
tensor product} of $A$ and $B$ we mean the model $(\E(A) \maxtensor \E(B), X(A) \maxtensor X(B))$, where the test space $X(A) \maxtensor X(B)$ is the maximal test space for $\E(A) \maxtensor \E(B)$.}
\end{definition}

These choices of these two test spaces are dictated by the desire to have the following 

\begin{proposition} If $AB$ is any locally tomographic composite of $A$ and $B$, then we have embeddings 
$A \mintensor B \rightarrow AB \rightarrow A \maxtensor B$. We also have, dually, 
$\Omega(A \maxtensor B) \leq \Omega(AB) \leq \Omega(A \mintensor B)$. \end{proposition} 
 
Thus, $A \mintensor B$ is the smallest possible locally tomographic composite of $A$ and $B$, in the sense of having the fewest possible effects. Dually, $\E(A \mintensor B)^{\ast} = \E(A)^{\ast} \maxtensor \E(B)^{\ast}$ has the {\em largest} possible state space among locally tomographic composites. One might say, roughly speaking, that $A \mintensor B$ admits {\em no} entanglement between 
effects, and, consequently, admits all possible entangled states. At the other extreme, $A \maxtensor B$ admits every possible entangled bipartite effect and, in consequence, admits no entanglement of states. 

If $\Omega(A)$ or $\Omega(B)$ is a simplex, then it is easy to show that $\V(A) \maxtensor \V(B) \simeq \V(A) \mintensor \V(B)$ and $\E(A) \maxtensor \E(B) \simeq \E(A) \mintensor \E(B)$. Thus, a classical system admits 
no entangled states or effects in any non-signaling composite with another system. There is a partial converse:
\\

\begin{theorem}[\cite{Namioka-Phelps}] The following are equivalent: 
\begin{itemize}
\item[(a)] $\Omega(A \maxtensor B)$ contains no entangled state for any 
model $B$, 
\item[(b)] $\Omega(A \maxtensor B)$ contains no entangled state, where $B$ is the square bit (Example....), 
\item[(c)] $\Omega(A)$ is a simplex. \\
\end{itemize}
\end{theorem} 

It follows that any {\em non}-classical system $A$ --- one with a non-simplicial state space -- will admit {\em some} locally tomographic, non-signaling composite $AB$ that admits entangled states. In this sense, entanglement is a highly generic phenomenon in non-classical 
probability theory. 

\subsection{Monoidal Probabilistic Theories} 

Earlier, we decided to represent a probabilistic {\em theory} as a category of probabilistic models with positive mappings 
as morphisms. It is not unreasonable to require that, if $A, B$ and $C$ are three systems, we should be able to form tripartite composites $(AB)C$ and $A(BC)$. We'd perhaps like to require that these be the same, 
i.e., that we have an {\em associative} rule of composition. This is not a trivial requirement ---  one can readily imagine situations in which the composition of systems might not be associative\footnote{Consider, for instance, the case of \[(\mbox{Farmer} \otimes \mbox{Hen} ) \otimes \mbox{Fox} \ \ \mbox{\em vs.} \ \ \mbox{Farmer} \otimes (\mbox{Hen} \otimes \mbox{Fox}).\]} --- but it is a natural one. 

A {\em symmetric monoidal category} is a category $\C$, equipped with a bi-functor $\C \times \C \stackrel{\otimes}{\longrightarrow} \C$, such that for all $A, B, C, D \in \C$, 
\[A \otimes (B \otimes C) \simeq (A \otimes B) \otimes C \ \ \text{and} \ \ A \otimes B \simeq B \otimes A\]
by means of natural isomorphsism $\alpha_{A;BC}$ and $\sigma_{A,B}$ belonging to $\C$; and also equipped with a {\em tensor unit}, $I$, and natural isomorphisms 
\[I \otimes A \simeq A \simeq A \otimes I\]
This point of view has been extensively developed in the the categorical semantics for quantum theory developed by Abramsky-Coecke and Selinger \cite{Abramsky-Coecke, CoeckeProcess, Selinger}, and also in the work of Baez and his students \cite{Baez, Baez-Stay}. 

\begin{definition} \label{def: monoidal theory}
A {\em monoidal probabilistic theory} is a probabilistic theory $\C$,
equipped with a rule of composition $A, B \mapsto AB$ assigning, to
each pair of models $A,B \in \C$, a composite $AB$ in the sense of
Definition \ref{def: composite}, and making $\C$ a symmetric monoidal
category. We shall say that $\C$ is {\em non-signaling}, respectively
{\em locally tomographic}, iff $AB$ is non-signaling or locally
tomographic for every pair $A, B \in \C$. \end{definition}

This definition implies that, for all $A, B \in \C$ and all states $\alpha \in \Omega(A)$, $\beta \in \Omega(B)$, there is a distinguished product state $\alpha \otimes \beta $ with $(\alpha \otimes \beta)(xy) = \alpha(x)\beta(y)$ 
for all $x \in X(A)$, $y \in X(B)$. Similarly, for any (dual) processes $\tau_1 \in \C(A)$ and $\tau_2 \in \C(B)$, there exists a process $(\tau_1 \otimes \tau_2) \in \C(AB)$ with 
$(\tau_1 \otimes \tau_2)(a \otimes b) = \tau_1(a) \otimes \tau_2(b)$ for all effects $a \in \E(A)$ and $b \in \E(B)$.

Finite-dimensional classical and quantum probability theory are both monoidal with respect to their usual rules of composition. The minimal and maximal tensor products are each naturally associative, and hence make the category of {\em all} probabilistic models into a monoidal probablistic theory; but neither is entirely satisfactory: the former provides for entangled states, but does not permit entangled effects, while the latter provides for entanglement between effects, but allows none between states. That a probabilistic theory support a {\em single} ``tensor product" that accommodates entanglement of both states and effects, is a non-trivial constraint. To be sure, one might consider probabilistic theories equipped with more than one rule of composition; however, the interactions among different non-signaling compositions on a given theory can be very delicate. 
It therefore seems reasonable to begin by investigating the simpler possibilities for a theory equipped with a single privileged, 
monoidal rule of composition. Accordingly, 
{\bf \emph{in the balance of this paper, we work in a monoidal probabilistic theory $\boldsymbol \C$.}}\\

\noindent{\bf Historical Remarks} Tensor products of compact convex sets or of 
order-unit spaces were studied in a number of papers in the late 1960s, notably that of Namioka and Phelps \cite{Namioka-Phelps}. The fact that the marginal of an entangled pure state must be a mixed state already appears there, albeit not in these terms, as do the definitions of what we are calling the maximal and minimal tensor products. Our treatment composite systems derives  from that of by Foulis and Randall \cite{Foulis-Randall81, KRF}. 
Some first attempts to understand probabilistic theories as symmetric monoidal categories of probabilistic models can be found in \cite{BW09, BDW10}; work in this direction is ongoing. \\



\section{Post-Classical Information Processing}

As we've seen, entangled bipartite states and effects arise very
naturally, not only in quantum theory, but in almost any context in
which we form non-signaling composites of non-classical systems. While
this observation goes back at least to \cite{KRF, Klay} in the late
1980s, it remained unexploited. Entanglement lies at the heart of
quantum information theory, so it natural to wonder to what extent
quantum information-theoretic results carry over to other
non-classical settings.  It turns out that a great many such results
do have analogues for probabilistic theories that are far more general
than quantum mechanics. While the exploration of this post-classical
information theory is still in its infancy, it has already shed
considerable light on the scope and meaning of several key
quantum-informational results.

In this section, we review in some detail two of these. The first is
the no-cloning theorem, and its generalization, the no-broadcasting
theorem. These hold in {\em any} finite-dimensional theory having a
state space that is not a simplex. The second is the existence of a
teleportation protocol, or, a bit more generally, of an
entanglement-swapping protocol. Here, some restrictions need to be
made, but they are of moderate strength.  For example, 
any monoidal probabilistic
theory in which individual systems are {\em weakly self-dual},
and composites include isomorphism states $\omega$ and effects $f$ 
corresponding to isomorphisms
$\hat{\omega}$, $\hat{f}$ witnessing the weak self-duality, supports
a certain kind of teleportation.  Moreover, when viewed in this
generality, teleportation loses most of its mystery: it is simply a
form of classical conditioning, one which appears startling only owing
to the appearance of isomorphism states.


\subsection{Cloning and broadcasting} 
To {\em clone} a state of a system $A$ means, very broadly, to produce
two \emph{independent} copies of that state by means of some physical
process. In the present formalism, if the initial state belongs to a
system $A$, this would require a positive linear mapping
\[\phi : \V(A) \rightarrow \V(AA)\]
such that $\phi(\alpha) = \alpha \otimes \alpha$. There is no difficulty producing such a mapping: indeed, the constant mapping $\Omega(A) \rightarrow \Omega(AA)$ given by 
$\beta \mapsto \alpha$ for all $\beta \in \Omega(A)$ is affine, and hence, extends uniquely to a positive linear mapping $\V(A) \rightarrow \V(AA)$. However, this mapping is 
(highly!) state-dependent. One might ask whether one could {\em jointly} clone a collection of states, say, $\alpha_1,...,\alpha_n$. That is: given such a set of states, can one 
find a {\em single}, norm-nonincreasing, positive linear mapping $\V(A) \rightarrow \V(AA)$ that clones them all, in the sense that $\phi(\alpha_i) = \alpha_i \otimes \alpha_i$ for all $i$? 

If the states $\alpha_i$ are jointly distinguishable, the answer is yes. If $\{a_i\}$ is an observable 
on $A$ with $\alpha_i(a_i) = 1$; then the mapping 
\[\phi(\beta) = \sum_{i} \beta(a_i)~ \alpha_{i} \otimes \alpha_{i}\]
does the trick. The {\em no-cloning theorem} is essentially the
converse: if there exists a single process that will clone all of the
states $\alpha_1,...,\alpha_n$, then there exists an observable that
distinguishes them. In the case of a discrete classical model, where
all pure states are jointly distinguishable, this is no restriction
on the clonability of pure states;
but quantum pure states, which are not jointly distinguishable, are in
general not jointly clonable. 

The quantum no-cloning theorem was first
proved, independently, by Wootters and Zurek \cite{Wootters-Zurek} and
by Dieks \cite{Dieks}. That the same result holds for arbitrary
probabilistic theories is proved in \cite{BBLW06}. We omit the proof
here, but the idea is simple: if we can clone each of the states
$\alpha_1,...,\alpha_n$ with a single mapping, then by iterating this
process, we can create arbitrarily large ensembles of independent
copies of an unknown state $\alpha \in \{\alpha_1,...,\alpha_n\}$ and,
by making measurements on this ensemble, we can use statistics to
distinguish among them.

We say that a state $\rho \in \Omega$ is {\em broadcast} by an
affine mapping $\phi : \Omega \rightarrow \Omega \otimes \Omega$ iff 
the bipartite state $\phi(\rho)$ has marginal states $\phi(\rho)_1$ and $\phi(\rho)_2$ both equal to 
$\rho$. If $\rho$ can be expressed as a mixture of distinguishable --- hence, clonable --- states 
$\alpha_1,...,\alpha_n$, say $\rho = \sum_{i} t_i \alpha_i$, then one can broadcast $\rho$ using 
a cloning map $\phi$ for the states $\alpha_1,...,\alpha_n$: 
the state $\phi(\rho) = \sum_{i} t_i  \alpha_i \otimes \alpha_i$ 
has both marginal states equal to $\rho$, as required. 
The quantum {\em no-broadcasting theorem} of Barnum et al. \cite{Barnumetal96}
tells us that, conversely, 
two {\em quantum} states are jointly broadcastable iff, regarded as
density operators, they commute --- which, by the Spectral Theorem, is equivalent to requiring 
that all are convex combinations of some single set of distinguishable pure states.  In fact, 
this is a corollary of a more general result:

\begin{theorem}[\cite{BBLW06, BBLW07}] Let $\Gamma$ be the set of states broadcast by an affine
mapping $\phi : \Omega \rightarrow \Omega \otimes \Omega$. Then
$\Gamma$ is the simplex generated by a set of
distinguishable states in $\Omega$, which are cloned by $\phi$. 
\end{theorem}

(Although we omit the proof here, it is not especially difficult. This is in contrast to earlier proofs of the quantum no-broadcasting result \cite{Barnumetal96, Lindblad}, which were not especially easy.)

\subsection{Remote Evaluation} Suppose $\C$ is a locally tomographic, monoidal probablilistic theory. Consider two parties, Alice and Bob, occupying arbitrarily distant sites. Suppose that Alice controls a pair of 
systems, say $A_o, A_1 \in \C$, while Bob controls a system $B \in \C$. Since $\C$ is monoidal, we can represent Alice's two systems together as a single bipartite system $A = A_o A_1$, 
and the entire Alice-Bob system, by the tripartite composite $AB = (A_o A_1)B \simeq A_o (A_1 B)$. 

Now suppose that the composite system $A_1 B$ is in a state $\omega$, while Alice's system $A_o$ is in a state 
$\alpha$, independent of the $A_1B$ sub-system. Then the total state of the system $AB = A_o (A_1 B)$ is $\alpha \otimes \omega$. 
Now let Alice make a measurement on her system $A = A_o A_1$, obtaining a result represented by an effect $f \in \E(A)$; suppose 
Bob also makes a measurement on his system, $B$, obtaining a result represented by an effect $b \in \E(B)$, 
so that the joint outcome of these two measurements is $f \otimes b$. 

\begin{lemma}[Remote Evaluation]  With notation as above, let $\widehat{\omega} : \E(A_1) \rightarrow \V(B)$ and 
$\widehat{f} : \V(A_o) \rightarrow \E(A_1)$ be the conditioning and co-conditioning maps associated with the 
state $\omega$ and the effect $f$. Then, for all $\alpha \in \V(A_o)$ and all $b \in \E(B)$, 
\begin{equation}
(\alpha \otimes \omega)(f \otimes b)  = \hat{f}(\hat{\omega}(\alpha))(b).\end{equation}
\end{lemma} 

The proof is easy: one simply checks that the formula is correct when $\omega$ is a product state and $f$ is a product effect. Since we are working with locally tomographic composites, product states and effects span $\E(A_1B)^{\ast}$ and $\E(A_o A_1)$, respectively, so (\theequation) holds for all choices of $\omega$ and $f$. 
Nevertheless, the result is somewhat surprising, for it asserts that the mapping 
\[\tau := \hat{\omega} \circ \hat{f} : \V(A_0) \rightarrow \V(B)\] 
can be implemented, probabilistically, by means of a preparation of $A_1B$ in the joint state $\omega$ and a (successful) observation of $f$ on $A_o A_1$. In particular, when Alice observes the effect $f$, the corresponding un-normalized conditional state of Bob's system is 
\[(\alpha \otimes \omega)(f \otimes - ) = \tau(\alpha).\]
Note that the probability of the process $\tau$ occurring in state $\alpha$ is $u_B(\tau(\alpha))$, which is 
is exactly the marginal probability $(\alpha \otimes \omega)_{1}(f)$ of Alice's obtaining $f$. 
In what follows, we refer to the pair $(f,\omega)$ as a {\em remote evalution} protocol for the process $\tau = \hat{f} \circ \hat{\omega}$.



We can reformulate the notion of conditioning and co-conditioning 
map, and the remote evaluation Lemma (Lemma 5),  in purely
categorical terms.  In fact, both make sense in any symmetric
monoidal category $\C$. Given objects $A, B \in C$ and a morphism $ \omega : A \otimes B \rightarrow I$, 
there is a canonical mapping
$\hat{\omega} : \C(I,A) \rightarrow \C(B,I)$ given by
\begin{equation}\label{eq:remote-eval-again-dual}
{\xymatrix@=12pt{ B \ar@{->}^{a \otimes \id_{B}}[rr] \ar@{->}_{\hat{\omega}(a)}[ddrr] & & A \otimes B \ar@{->}[dd]^{\omega}\\
& & \\
& & I }}.
\end{equation}
Dually, if $f  \in \C(I,A \otimes B)$, there is a
natural mapping $\hat{f} : \C(A,I) \rightarrow \C(I,B)$ given by
\begin{equation}\label{eq:remote-eval-again}
{\xymatrix@=12pt{ I \ar@{->}^{f}[rr] \ar@{->}_{\hat{f}(\alpha)}[ddrr] & & A \otimes B \ar@{->}[dd]^{\alpha \otimes \id_{B}}\\
& & \\
& & B }}
\end{equation}
If $\C$ is a monoidal probabilistic theory, then $\hat{\omega}$ and
$\hat{f}$, defined in this way, correspond exactly to the conditioning
and co-conditioning maps associated with the bipartite state $\omega :
A \otimes B \rightarrow I$ and effect $f : I \rightarrow A \otimes B$.
Combining diagrams (3) and (4), and taking advantage of the monoidal
structure of $\C$ --- in particular, the fact that $\alpha \otimes
\omega = (I \otimes \omega) \circ (\alpha \otimes \id_{A_o \otimes
  A_1})$ --- we have
\begin{equation} \hat{\omega}(\hat{f}(\alpha) \otimes \id_{B}) = \omega \circ (\alpha \otimes \id_{A_1B}) \circ (f \circ \id_{B}) = (\alpha \otimes \omega) \circ (f \otimes \id_B)\end{equation}
which precisely expresses Lemma 5. 
\begin{equation}\label{eq:remote-eval-3}
{\xymatrix@=12pt{ 
 & & & A_o \otimes A_1 \otimes B \ar@{->}^{\alpha \otimes \id_{A_1B}}[dd]\\
 & & & \\
B = I \otimes B \ar@{->}^{f \otimes B}[rrruu] \ar@{->}^{\hat{f}(\alpha) \otimes \id_{B}}[rrr] \ar@{->}_{\hat{\omega}(\hat{f}(\alpha))}[rrrdd]
& & & A_1 \otimes B \ar@{->}[dd]^{\omega}\\
& & & \\
& & & I}}
\end{equation}

This has an important corollary. Since $\omega \circ (\alpha \otimes \id_{A_1 B}) = \alpha \otimes (\id_{A_o} \circ \omega)$, 
we can re-write (\theequation) as 
\[\hat{\omega}(\hat{f}(\alpha) \otimes \id_{B}) = \alpha \circ (\id_{A_o} \otimes \omega) \circ (f \otimes \id_{B})\]
Thus, the dual process $\tau^{\ast} : \E(B) \rightarrow \E(A)$ corresponding to the 
process $\tau = \hat{\omega} \circ \hat{f}$ arising in the 
remote evaluation protocol, is in fact a morphism in $\C(A_o,B)$. \\

%


\noindent{\bf Conclusive Teleportation} In the special case in which
the models $A_o, A_1$ and $B$ are isomorphic and weakly self-dual, we
can consider a remote evaluation protocol in which both the effect $f
\in \E(A)$ and the state $\omega \in \Omega(A_1B)$ correspond to order
isomorphisms $\hat{f} : \V(E_o) \simeq \E(A_1)$ and $\hat{\omega} :
\E(A_1) \simeq \V(B)$.  In this case, the process $\tau
= \hat{\omega}\circ \hat{f}$ is again an order-isomorphism.
If this scenario is repeated many times, Bob can perform sufficiently
many measurements to determine $\tau(\alpha)$ with reasonable
confidence, and then {\em compute} the value
of $\alpha$. On the other hand, if $\tau$ is probabilistically
reversible, in a single run of the scenario Bob can actually correct
his state, with non-zero probability, so that it agrees with
$\alpha$.  In this case, we may say that the state $\alpha$ has been
{\em teleported} from Alice's system $A_o$ to Bob's system $B$, and
refer to $(f,\omega)$ as a {\em teleportation protocol}. If $\tau$ is 
reversible with probability $1$, we shall say
that $(f,\omega)$ is a {\em strong} teleportation protocol. \\



\noindent{\bf Deterministic Teleportation} Suppose now that Alice has
access to an observable $\{f_i\}$ on $A = A_oA_1$, with each of the
effects $f_i$ an isomorphism effect. Each of these effects, in
combination with the isomorphism state $\omega$, gives rise to a
conclusive teleportation protocol, implementing the order-isomorphism
$\tau_i = \hat{\omega} \circ \hat{f}_i : \V(A_o) \simeq \B(B)$. If
Alice is permitted to communicate (classically) with Bob, then upon
observing outcome $f_i$, she can instruct Bob to implement the inverse
process $\tau^{-1}$, which he can do with probability $c_i :=
u_{B}\tau_{i}^{-1}(\alpha)$. It follows that the post-measurement
state of Bob's system will be $\sum_{i} c_i \alpha = \alpha$. 
particular, $\sum_i c_i = \sum_i u_B \tau_{i}^{-1}(\alpha)$. 
Say that $A$ {\em supports a deterministic teleportation protocol} iff
there exists such an observable $\{f_i\}$ and such a state
$\omega$. 

\begin{theorem}[\cite{BBLW08}] Suppose there exist a finite group $G$ acting transitively on 
$A$'s pure states, and a $G$-equivariant order-isomorphism $\E(A)
  \simeq \E(A)^{\ast}$. Then $A$ supports a deterministic
  teleportation protocol.\end{theorem}


\noindent{\bf Entanglement Swapping} Suppose that, like Alice, Bob controls a bipartite system $B = B_1 B_2$. Assume here 
that $A_o, A_1, B_1$ and $B_0$ are all isomorphic to one another. Given an entangled state $\omega$ between $A_1$ and $B_1$, 
and isomorphism effects $f$ on $A = A_o A_1$ and $g$ on $B = B_1 B_0$, we find that, for any state $\mu$ on $A_0 A_0$, we have 
(up to the obvious symmetrizers and associators) 
\[(f \otimes g)(\mu \otimes \omega) = g(\hat{\omega} \circ \hat{f} \circ \hat{\mu}^{\ast}).\]
Since this holds for any choice of $g \in \E(B)$, we have 
\[(\mu \otimes \omega)_{B|f} = \hat{\omega} \circ \hat{f} \circ \hat{\mu}^{\ast}\]
If $\tau = \hat{\omega} \circ \hat{f}$ is probabilistically reversible, 
then upon Bob's executing the reverse process, the state
$\mu$ has been transferred from $A_0 B_2$ to $B = B_1
B_0$. \\ 

\noindent{\bf Teleportation and Compact Closure}  Let $\C$ be any symmetric monoidal category. A {\em dual} for an object $A \in \C$ is an object $B \in \C$, together with two morphisms,
$\eta : I \rightarrow B \otimes A$ and 
$\epsilon : A \otimes B \rightarrow I$ --- called the {\em unit} and {\em co-unit}, respectively --- such that
\begin{equation} 
(\epsilon \otimes \id_{A}) \circ (\id_{A} \otimes \eta) = \id_{A} \ \ \text{and} \ \ (\id_{B} \otimes \epsilon) \circ (\eta \otimes \id_{A}) = \id_{B}
\end{equation}
In view of the discussion above, if $\C$ is a monoidal probabilistic theory and $f, \omega$ is a conclusive teleportation protocol for a pair of systems $A, B \in \C$, then the remote evaluation lemma tells us that 
$f$ and $\omega$ function as a unit and co-unit, respectively, for $A$ and $B$.
A symmetric monoidal category in which every object has a dual is said to 
be {\em compact closed}. A {\em compact structure} on 
a compact closed category is a specification, for every object $A \in \C$, of a distinguished dual $A' \in \C$. Where $A = A'$ for every $A \in \C$, this 
structure is {\em degenerate}.\footnote{Duals, where they exist, are canonically isomorphic. Hence, for most purposes, the 
choice of one rather than another object as ``the" dual is irrelevant. The existence of a degenerate compact structure is, however, 
a real constraint \cite{BDW10, Selinger10}.}

\begin{proposition}[\cite{BDW10}] Let $\C$ be a monoidal probabilstic theory. The following are equivalent.
\begin{itemize}
\item[(a)] $\C$ admits a compact closed structure.
\item[(b)] Every $A \in \C$ can be teleported through some $B \in \C$;
\item[(c)] Every morphism in $\C$ has the form $\hat{\omega} \circ \hat{f}$ for some bipartite state $\omega$
and bipartite effect $f$ in $\C$. 
\end{itemize}
\end{proposition}

\noindent{\em Proof:} The equivalence of (a) and (b) is clear from the
preceding discussion. To see that these are in turn equivalent to (c),
suppose first that (a) and (b) hold.  Choose for each $A \in \C$ a
dual system $A'$, a state $\omega_A \in \C(A \otimes A', I)$, and an
effect $f_A \in \C(I,A' \otimes A)$ with $\hat{\omega_{A}} =
\hat{f_{A}}^{-1}$. Then for any morphism $\tau \in \C(A,B)$, let
$f_{\tau} \in \C(I,A \otimes B)$ be the effect $f_{A} \circ (A'
\otimes \tau)$. It is easily checked that then $\hat{f}_{\tau} = \tau
\circ \hat{f_{A}}$, so that $\tau = \hat{f}_{\tau} \circ
\hat{\omega_{A}}$.  Conversely, if (c) holds, then for each $A$, the
identity mapping $\id_{A}$ factors as $\hat{\omega}_{A} \circ
\hat{f}_{A}$ for some $\omega_{A} \in \C(B \otimes A,I)$ and some $f
\in A \otimes B$. It follows that $\hat{\omega}_{A} =
\hat{f}_{A}^{-1}$, so this gives us a compact closed
structure. $\Box$\\




\subsection{Steering}

Let $B$ be a probabilistic model. An  {\em ensemble} for a state $\beta \in \Omega(B)$ is a 
finite set of of states $\beta_i \in \V(B)_+$ such that $\sum_i \beta_i = \beta$. We can 
understand such an ensemble as representing one possible way of {\em preparing} the state $\beta$, 
namely, to choose one of the normalized states $\hat{\beta}_i := \beta_i / u(\beta_i)$ with 
probability $p_i = u_B(\beta_i)$. 

One way to do {\em this} is to begin with a bipartite state $\omega$
on a non-signaling composite $AB$, with marginal $\omega_2 =
\beta$. Then for any observable $E = \{a_i\}$ on $A$, the
un-normalized conditional states $\beta_i := \hat{\omega}(a_i)$ are an
ensemble for $\beta$. That is: by measuring $E$, we prepare not only
the marginal state $\omega_B$, but {\em a particular ensemble} for
this state. By choosing to measure a different observable, we will
typically obtain a different ensemble for $\beta$.  If $A$ and $B$ are
{\em quantum} systems, and if $\omega$ is a pure entangled state of
$AB$, then {\em any} ensemble for $\omega_2$ can be obtained in this
way from a suitable choice of measurement on $A$.  This phenomenon was
first observed by Schr\"odinger \cite{Schroedinger}, who called it
{\em steering}.  The concept extends readily to the setting of an
arbitrary non-signaling composite.

\begin{definition}{\em  Let $AB$ be a non-signaling composite of probabilistic 
models $A$ and $B$. A bipartite state $\omega \in AB$ 
is {\em steering for its $B$ marginal}, or {\em $B$-steering}, for short, iff, for every ensemble (convex
decomposition) $\omega_{2} = \sum_i \beta_i$, where $\beta_i$ are
un-normalized states of $B$, there exists an observable $E = \{a_i\}$
on $A$ with $\beta_i = \hat{\omega}(a_i)$. We say that $\omega$ is
{\em bi-steering} iff it's steering for both marginals.}
\end{definition}



The relevance of steering to information processing became evident when Bennett and
Brassard \cite{BB84}, in the same paper that introduced quantum key
distribution, considered a natural quantum scheme for another
important cryptographic primitive, bit commitment, and showed that
ensemble steering can be used to break it.  In the proposed scheme, the two
possible values to which Alice can commit  are represented by two distinct
ensembles for the same density matrix. She is to send samples from the
ensemble to Bob in order to commit, and later reveal which states she
drew so that Bob can check that she used the claimed ensemble.
However, by sending to Bob, not a draw from the ensemble, but one of two 
systems in an entangled 
pure bipartite state with the specified density matrix as its marginal. 
Keeping the other system, she can realize either ensemble  after she has
already sent the systems to Bob by making measurements on her
entangled system, enabling her to perfectly mimic commitment to either
bit.  

Later Mayers, and Lo and Chau, showed that \emph{no}
information-theoretically secure quantum bit commitment protocol can
exist. The techniques they used to defeat putative protocols do not
literally use steering, but are closely related to the
Bennett-Brassard steering attack, in particular in Alice's retention
of a system \emph{purifiying} the systems she sends to Bob in the
course of the protocol.

The paper \cite{BGW09} studies steering in the context of general probabilistic 
theories.  If $\alpha$ is any state on $A$ and $\beta$ is a {\em pure} state on
$B$, then $\omega = \alpha \otimes \beta$ is trivially steering for
$\omega_{2} = \beta$ since the latter has no non-trivial ensembles. In particular, 
any pure product state will be steering for both of its marginals. 
Any isomorphism 
state $\omega \in \V(AB)$ will also be steering. 


It follows almost immediately from the definition, that if $\omega$ is
steering for its $B$-marginal, then the image, $\homega(\E(A)_{+})$, of 
the positive cone in $\E(A)$, is a face of
$\V(B)_{+}$. Indeed, we have

\begin{lemma} \label{lemma: steering face}
If $\omega$ is steering, then $\homega(\E(A)_+) = \face(\omega_{2})$. \end{lemma}

Here $\face(\omega_2)$ refers to the face generated by $\omega_2$, i.e, the smallest 
face of $\V(B)_+$ containing $\omega_2$. The converse of Lemma (\thelemma) is false. 

A probabilistic theory $\C$ \emph{supports uniform universal steering}
if, for every system $B \in \C$, there exists a system $A_B \in \C$
such that every state $\beta \in A$ is the marginal of some $B$-steering state $\omega 
\in A_B B$.  If one can always take $A_B = A$, we say that $\C$  {\em supports universal self-steering}. 




\begin{proposition} \label{cor: injective steering pure}  
Let $\omega \in \Omega(AB)$ be steering for $\omega_2$, where $\omega_2$ is interior
to $\V(B)_{+}$, so that $\face(\omega_2) = \V(B)_+$. If $\homega$ is
injective (non-singular), then $\homega$ is an order isomorphism.  If
$\V(B)$ is irreducible, therefore, by
Proposition \ref{prop: isomorphism states pure}, $\homega$ it is pure. 
\end{proposition}

In other words, if $A$ and $B$ have the same dimension, then the
states that are steering for an {\em interior marginal} are precisely the
isomorphism states (and hence, are steering for both marginals).

Steering is closely related to an important property of quantum theory called 
\emph{homogeneity}.  

\begin{definition}{\em Let $\G$ be a group of order-automorphisms of an ordered vector space $\E$. We say that $\E$ is {\em homogeneous with respect to} $\G$ if $\G$ acts transitively on the {\em interior} of the positive 
cone $\E_+$. That is, for every pair of interior points $a, b$ of
$\E_+$, there exists an element $g \in \G$ with $ga =
b$.  We say $\E$ is \emph{homogeneous} if it is homogeneous with respect to 
some group of order-automorphisms, or, equivalently, if it is homogeneous with respect to the group $\Aut(\E)$ of all order-automorphisms. }\end{definition}

It can be shown that the cone $\L_{+}(\H)$ of positive operators on a finite-dimensional Hilbert space $\H$ is homogeneous 
with respect to the group of order-automorphisms of $\L(\H)$. As we discuss below in Section 5, the combination of 
homogeneity and strong self-duality comes close to characterizing finite-dimensional quantum theory among 
probabilistic theories generally. More precisely, the {\em Koecher-Vinberg Theorem} asserts that if $\E$ is an ordered linear space whose positive cone $\E_+$ is both homogeneous and self-dual, then $\E$ can be given the structure of a euclidean Jordan algebra.  With this in mind, the following result is particularly intriguing:

\begin{theorem} \label{theorem: isomorphism homogeneity} 

For a model with irreducible state space $\V(A)$ the following are equivalent:\\
\noindent
(a) $A$ is homogeneous;\\
(b) Every normalized state in the interior of $\Omega(A)$ is the
$A$-marginal of an isomorphism state in $B \maxtensor A$, where $B$ is any 
(fixed) model with state space order-isomorphic to $\V(A)^{\ast}$. 
\end{theorem}

From this we obtain:

\begin{corollary} \label{cor: isomorphism homogeneity wsd} 

For any model with 
irreducible state space $A$, 
the following are equivalent:\\
\noindent
(a) $\V(A)_+$ is weakly self-dual and homogeneous;\\
(b) Every normalized state in the interior of $\Omega(A)$ is the
  marginal of an isomorphism state in $A \maxtensor A$.
\end{corollary}

Corollary \ref{cor: injective steering pure}, combined with Theorem
\ref{theorem: isomorphism homogeneity}, gives

\begin{proposition}
In any theory that supports universal uniform steering, every
irreducible, finite-dimensional state space in the theory is
homogeneous.
\end{proposition}

In light of Corollary \ref{cor: isomorphism homogeneity wsd}, we also have
\begin{proposition}
In any theory that supports universal self-steering, every irreducible,
finite-dimensional state space in the theory is homogeneous and weakly
self-dual.
\end{proposition}

Therefore, the distance between probabilistic theories allowing
universal self-steering, and those whose state-spaces are
Jordan-algebraic is just that between weak and strong self-duality.

In \cite{BDLT} it was shown that an asymptotically exponentially
secure bit commitment protocol, based (like the original
Bennett-Brassard one-qubit protocol) on the nonuniqueness of convex
decomposition in nonclassical state spaces, exists in any theory 
containing some nonclassical state spaces, coupled only by the
minimal tensor product (so that there is no entanglement between them).  In a
nonclassical theory in which all states can be steered, by contrast,
this type of bit commitment protocol can always be defeated.


\subsection{Entropy and Information Causality}

Classical information theory begins with the Gibbs-Shannon entropy
$H(p) = -\sum_i p_i \log(p_i)$ of a discrete probabiilty weight
$p_1,...,p_n$. Analogously, in quantum theory the {\em von Neumann
  entropy} of the state corresponding to a density operator $\rho$ is
given by $S(\rho) := \tr \rho \log{\rho}$.
This is related to the classical Gibbs-Shannon entropy in two
important ways. On one hand, $S(\rho)$ is the minimum of the
Gibbs-Shannon entropies $-\sum_i p_i \log{p_i}$ of the probability
weights $p_i = \tr(\rho e_i)$ that $\rho$ induces on quantum tests
$\{e_i\}$. (This turns out to be achieved when the measurement is in a
diagonalizing basis). Alternatively, $S(\rho)$ is the minimum
Gibbs-Shannon entropy of the probabilities $p_i$ arising in
representations of $\rho$ as a mixture $\rho = \sum_i p_i \rho_i$ of
pure states $\rho_i$. (This again turns out to be achieved for an
ensemble whose states are the rank-one projectors corresponding to a
diagonalizing basis).

Both of these characterizations make sense in the context of an
arbitrary probabilistic model, but in general, they are not
equivalent.


\begin{definition} Let $\alpha$ be a state on $A$. For each test $E \in \A(A)$, 
define the {\em local measurement entropy} of $\alpha$
at $E$, $H_{E}(\alpha)$, to be the classical (Shannon) entropy of
$\alpha|_{E}$, i.e.,
\[H_{E}(\alpha) := - \sum_{x \in E} \alpha(x) \log(\alpha(x)).\]
The {\em measurement entropy} of $\alpha$, $H(\alpha)$, is the
infimum of $H_{E}(\alpha)$ as $E$ ranges over $\A(A)$, 
i.e.,
\[H(\alpha) := \inf_{E \in \A(A)} H_{E}(\alpha).\]
\end{definition}

 Note that the measurement entropy of a state $\alpha \in \Omega(A)$
 depends entirely on the structure of the test space $\M(A)$, and
 not on the geometry of the state space $\Omega$.

We shall assume in what follows that the measurement entropy of a
state is actually achieved on some test, i.e., that $H(\alpha) =
H_{E}(\alpha)$ for some $E \in \A(A)$. This is the case in quantum
theory, and can be shown to hold much more generally, given some
rather weak analytic requirements on the model $A$ (\cite{Entropy},
Appendix B.) It follows that $H(\alpha)=0$ if and only if there is a
test such that $\alpha$ assigns probability $1$ to one of its
outcomes.\\

\noindent{\em Notation:} It will often be
convenient to write $H(\alpha)$ as $H(A)$, where context makes clear
which state is being considered. 
If $AB$ is a non-signaling composite, and $H(AB)$ reprents $H(\omega)$, we shall write 
$H(A)$ and $H(B)$ for the marginal entropies $H(\omega_1)$ and $H(\omega_2)$. It is easily checked 
that the measurement entropy is {\em subadditive}, i.e., 
\[H(AB) \le H(A) + H(B).\]

\begin{definition} Let $\alpha$ be a state on $A$. The {\em
mixing} (or {\em preparation}) entropy for $\alpha$, denoted
$S(\alpha)$, is the infimum of the classical (Shannon) entropy
$H(p_1,...,p_n)$ over all finite convex decompositions $\alpha =
\sum_i p_i \alpha_i$ with $\alpha_i$ pure states in $\Omega(A)$. 
\end{definition}

Again, we write $S(A)$ for $S(\alpha)$ where $\alpha$ belongs to the
state space $\Omega$ of a system $A = (\A,\Omega)$.  In
contrast to measurement entropy, the mixing entropy of a state depends
only on the geometry of the state space $\Omega$, and is independent
of the choice of test space $\A(A)$.  The mixing entropy is
essentially the same as the entropy defined for elements of compact
convex sets by A. Uhlmann in \cite{Uhlmann70}. 

We call a theory \emph{monoentropic} if mixing entropy equals
measurement entropy, for every state of every model in the theory.
Appendix B of \cite{Entropy} considers 
some implications of
monoentropicity. For instance, it is 
shown that any monoentropic model $A$ in which the set of pure states 
is closed in $\Omega(A)$ is sharp. 

We define conditional and mutual information in
terms of measurement entropy via formulas that also hold classically:

\begin{definition}
The \emph{conditional measurement entropy}
between $A$ and $B$ is defined to be
\begin{equation}\label{condmmtentropy}
H(A|B) := H(AB) - H(B).
\end{equation}
The [measurement-based] \emph{mutual information} is defined to be:
\begin{equation} \label{mutualinfdefgeneral}
I(A:B) := H(A) + H(B) - H(AB).
\end{equation}
\end{definition}

Intuitively, one might expect that $I(A:B)$
should not {\em decrease} if we recognize that $B$
is a part of some larger composite system $BC$ -- i.e., we might
expect that $I(A:B)
\leq I(A:BC)$. Simple algebraic manipulations (using
Eqs.~(\ref{condmmtentropy}) and (\ref{mutualinfdefgeneral}))
allow us to reformulate this condition in various ways.
\begin{lemma}\label{strongsublemma}
The following are equivalent:
\begin{itemize}
\item[(a)] $I(A:BC) \geq I(A:B)$
\item[(b)] $H(A|BC) \leq H(A|B)$
\item[(c)] $H(AB) + H(BC) - H(B) \leq H(ABC)$
\item[(d)] $I(A:B|C) \geq 0$, where $I(A:B|C) = H(A|C) + H(B|C) - H(AB|C)$.
\end{itemize}
\end{lemma}
The measurement entropy is said to be \emph{strongly subadditive} if
it satisfies the equivalent conditions (a)-(d).  
(Condition (c) is what is usually termed ``strong subadditivity'' (SSA).)  A
probabilistic theory in which conditions (a)-(d) are satisfied for all
systems $A, B$ and $C$ will also be called \emph{strongly
  subadditive}.  Despite the intution mentioned above, 
strong subadditivity can fail in general theories, which is perhaps a signal
that mutual information as defined above should not be interpreted in general
as ``the information each system contains about the other''.\\

\noindent{\bf The Holevo Bound and the Data Processing Inequality} The strong subadditivity inequality is crucial to deriving bounds on
many quantum information-transmission protocols, and the conditions
under which it is satisfied with equality are also of great
importance. Another extremely important inequality -- derivable, in the quantum
setting, from strong subadditivity -- is the \emph{Holevo bound}, which
figures in an expresssion for the highest achievable rate of classical
information transmission through a noisy quantum channel.

The standard formulation of the Holevo bound can apply to a general
theory, if the entropies are interpreted as measurement entropies: it
asserts that if Alice prepares a state $\rho = \sum_{x \in E} p_x
\rho_x$ for Bob, then, for any measurement $F$ that Bob can make on
his system,
\[
I(E:F) \leq \chi,
\]
where $\chi := H(\rho) - \sum_{x \in E} p_x H(\rho_x)$ (often called the {\em Holevo quantity}).

Suppose that Alice has a classical system $A =
(\{E\},\Delta(E))$ and Bob a general system $B$. Alice's system is to
serve as a record of which state of $B$ she prepared.  The
situation above is modeled by the joint state $\omega^{AB} = \sum_{x
  \in E} p_x \delta_{x} \otimes \beta_x$, where $\delta_x$ is a
deterministic state of Alice's system with $\delta_x(x) = 1$. Bob's
marginal state is $\omega_{2} = \sum_{x \in E} p_x \beta_x$. By
Lemma~\ref{goodgollymissmolly}, $H(\omega^{AB}) = H(A) + \sum_{x \in
  E} p_x H(\beta_x)$. Hence,
\begin{eqnarray*}
I(A:B) & = & H(A) + H(B) - H(AB) \\
& = & H(A) + H(B) - \left( H(A) + \sum_{x \in E} p_x H(\beta_x) \right) \\
& = & H(\omega_B) - \sum_{x \in E} p_x H(\beta_x) = \chi.
\end{eqnarray*}
So the content of the Holevo bound is simply that the mutual
information between the measurement of Alice's classical system and
any measurement on Bob's system is no greater than $I(A:B)$,
\[
I(E:F) \leq I(A:B).
\]
(While this is certainly natural, in general theories it does not always hold.)

Both strong subadditivity and the Holevo bound are instances of a more
basic principle.  The {\em data processing inequality} (DPI) asserts
that, for any systems $A$, $B$ and $C$, and any physical process ${\cal E} :
B \rightarrow C$,
\[
I(A:{\cal E}(B)) \leq I(A:B)\]
where $I(A:{\mathcal E}(B))$ refers to the mutual information of the state resulting 
from applying $\id_{A} \otimes {\cal E})$ to the state of $AB$.  
The strong subadditivity of entropy amounts to the DPI for the process
that simply discards a system (the \emph{marginalization map} $BC
\rightarrow C$). The Holevo bound is the DPI for the special case of
measurements, which can be understood as processes taking a system
into a classical system which records the outcome.\\

\noindent{\bf Information Causality} In a widely discussed paper
\cite{InfCausality}, , M. Pawlowski {\em et al.} introduced a constraint on a
non-signaling probabilistic theory, which they called {\em information
  causality}, in terms of the following protocol. Two parties, Alice
and Bob, share a joint non-signaling state, known to both of
them. Alice receives a random bit string $\e$ of length $N$; after
making measurements, she sends Bob message, $\f$, a bit-string of
length length $m$ or less.  Bob receives a radom variable $G$,
encoding a number, $k = 1,..., N$, which he takes as the instruction
to measure Alice's $k$-th bit. After making a suitable measurement,
and taking into account both its outcome and Alice's message, Bob
produces his guess, $b_k$.  Information causality is the requirement
that
\begin{equation}\label{ic}
\sum_{k=1}^{N} I(e_k : b_k | G = k) \leq m.
\end{equation} 
The main result of \cite{InfCausality} is that if a theory contains states
that violate the CHSH inequality by more than the
Tsirel'son bound, then it violates information
causality. In particular, if Alice and Bob can share PR boxes, then
using a protocol due to van Dam \cite{vanDam}, they can violate
information causality maximally, meaning that Bob's guess is correct
with certainty, and the left hand side of Equation~(\ref{ic}) is
$N$. Pawlowski \emph{et al.} also give a proof, using fairly standard
manipulations of quantum mutual information, that quantum theory {\em
  does} satisfy information causality.

One of the principle results of \cite{Entropy} is a suficient condition for 
a general probabilistic theory to be information-causal. The 
following is a strengthening of that result:  

\begin{theorem} \label{ictheorem}
Suppose that a theory is strongly subadditive, and satisfies the
Holevo bound.  Then the theory satisfies information causality. It
follows that any theory satisfying these conditions cannot violate
Tsirel'son's bound.
\end{theorem}

Since strong subadditivity and the Holevo bound follow from the data processing
inequality, we have the following:

\begin{corollary}
Any theory in which measurement-based mutual information 
satisfies the data processing inequality satisfies information 
causality.
\end{corollary}

In \cite{Entropy}, monoentropicity was assumed in addition to SSA and Holevo.
As noted there, it was only used to derive that $H(A|B) \ge 0$ when $A$ 
is classical.  However, this follows easily from strong subadditivity in the 
equivalent (cf. Lemma \ref{strongsublemma}) form $I(A:B|C) \ge 0$, when
we let $A$ and $B$ be identical perfectly correlated classical systems.  
We have 
\beqa
I(A:B|C) & = & H(A|C) + H(B|C) - H(AB|C) \\
&=& H(AC) - H(C) + H(BC) - H(C) - H(ABC) + H(C) \\
&=& H(AC) + H(BC) - H(ABC) - H(C). \\
\eeqa
Since $A,B$ are perfectly correlated classical systems, 
$H(AC) = H(BC) = H(ABC)$.  Consequently, in this case
$I(A:B|C) = H(AC) - H(C) \equiv H(A|C)$.  By SSA, this is $\ge 0$.\footnote{
The realization that Theorem 4 of \cite{Entropy} could be
strengthened this way grew out of discussions between some of 
the authors of \cite{Entropy} while the article was in press, but
too late for inclusion in the published version.}

\subsection{Other developments}  
There is much more to say about information processing in general
probabilistic theories than we have room to discuss here.  
We remark in particular on \cite{Barrett-Leifer}, in which a version
of the deFinetti theorem is proved for states on test spaces.\\


\section{Characterizing Quantum Theory}\label{sec: char}

As we've seen, a great number of information-processing phenomena
first discovered in association with quantum theory, are actually
rather more generally {\em post-classical}, rather than specfically
quantum-mechanical, in character. This brings us back to the question
of how to {\em characterize} quantum theory in operational or
probabilistic terms. The idea is to identify one or more
features of quantum theory that can be expressed in purely
operational-probabilistic terms --- roughly, without any special
reference to the Hilbert space structure, but only
in terms of primitive concepts such as states, effects, tests,
processes, etc. --- and that, taken together, {\em uniquely} specify
quantum (or quantum-plus-classical) models. This is an old problem,
and also a somewhat vague one, since what counts as a satisfactory
solution will be, to some extent, a matter of taste.  
Even so, striking progress has been made in the past several years, 
leading to several different, more-or-less satisfactory characgterizations of 
quantum mechanics as a probability theory [refs].
have been found. In this section, we review one of
these \cite{BW09, Wilce12, Wilce11, BW12}, which makes use of the
equivalence between homogeneous self-dual cones and Euclidean Jordan
algebras.

\subsection{Homogeneity and Self-Duality}

Let $\E$ be (for the moment) any finite-dimensional ordered linear space. Given a bilinear form $\Bi : \E \times \E \rightarrow \R$, we define the {\em internal dual} (with respect to $\Bi$) of 
the cone $\E_+$ to be the cone 
\[\E^{+} := \{ a \in \E | \forall x \in \E_+, \ \Bi(a,x) \geq 0\}.\]
We say that $\Bi$ is {\em positive on} $\E_+$, or simply {\em positive}, iff $\E_+ \subseteq \E^{+}$ --- in other words, if the linear mapping $\beta : \E \rightarrow \E^{\ast}$ given by 
$\beta(a)(x) = \Bi(a,x)$ is positive. 

\begin{definition}{\em $\E$ is {\em self-dual with respect to $\Bi$} iff $\E^{+} = \E_+$. We shall say that $\E$ is {\em weakly self-dual} iff there exists a bilinear form $\Bi$ with respect to 
which $\E$ is self-dual, and {\em strongly} self-dual, if there exists an {\em inner product} on $\E$ having this feature. }\end{definition}

Weak self-duality is equivalent to the existence of an isomorphism
state in $A \maxtensor A$. As discussed above, this is equivalent to
the requirement that there exist {\em some} composite of three copies of $A$
that supports a teleportation protocol, and to the requirement that
states on $A$ arise as marginals of steering states in a composite of
$A$ with itself \cite{BGW09}.  Strong self-duality is much less easy
to motivate, but we will discuss several ways in which it can be
justified in the next section.

Recall that $\E$ is {\em homogeneous} with respect to 
a group $\G$ of order-automorphisms if $\G$ acts transitively on the {\em interior} of the positive 
cone $\E_+$, so that for every pair of interior points $a, b$ of $\E_+$,  there exists an element $g \in \G$ with $ga = b$. 


Classical and quantum probabilistic models are both homogeneous and
self-dual. 
Somewhat more generally, let $\E$ be a
euclidean Jordan algebra. This is a finite-dimensional real
vector space $\E$ equipped with a commutative bilinar operation
$\bullet$ satisfying the {\em Jordan identity} $a^{2} \bullet (b
\bullet a) = (a^{2} \bullet b) \bullet a$ for all $a, b \in \E$, and
equipped with a canonical trace such that $\langle a, b \rangle :=
\tr(a \bullet b)$ is an innner product, with $\langle a \bullet b, c
\rangle = \langle a , b \bullet c\rangle$ for all $a, b, c \in
\E$. The set $\E_+ = \{ a^2 | a \in \E\}$ (where $a^2 = a \bullet a$)
is a cone in $\E_+$, and one can show is homogeneous with respect to the 
group of order-automorphisms of $\E$, and
self-dual with respect to the tracial inner product.  Remarkably,
there is a converse, to be found in work of M.  Koecher \cite{Koecher}
and E. Vinberg \cite{Vinberg}

If $G$ be any closed subgroup of $\Aut(\E)$, acting transitively on
the interior of $\E_+$, then $G$ is a Lie subgroup of $GL(\E)$. Let
$\g$ denote its Lie algebra, and let $\g_u$ denote the Lie algebra of
the stabilizer $G_u \leq G$ of the order-unit. The following
formulation of the Koecher-Vinberg Theorem summarizes the construction
of the Jordan product on $\E$. See \cite{Faraut-Koranyi} for a proof
(also, the Appendix to \cite{BW12} contains a fairly detailed outline
of the proof and some additional remarks pertinent to the precise
version given above):

\begin{theorem}[Koecher-Vinberg] Let $\E_+$ be self-dual with respect to some inner product on $\E$, and let 
$G$ be a closed, connected subgroup of $\Aut(\E)$, acting transitively on the interior of $\E_{+}$. Then 
\begin{itemize} 
\item[(a)] It is possible to choose a self-dualizing inner product on
$\E_+$ in such a way that $G_u = G \cap \O(\E)$ (where $\O(\E)$ is the 
orthogonal group with respect to the inner product);
\item[(b)] If $G = G^{\dagger}$ with respect to this inner product, then
$\g_u = \{ X \in \g | X^{\dagger} = -X\} = \{ X \in \g | Xu = 0\}$, and $\g = \g_u \oplus \p$, where
$\p = \{ X \in \g | X^{\dagger} = X\}$;
\item[(c)] In this case the mapping $\p \rightarrow \E$, given by $X
\mapsto Xu$, is an isomorphism.  Letting $L_a$ be the unique element
of $\p$ with $L_a u = a$, define
\[a \bullet b = L_{a} b\] 
for all $a, b \in \E$. Then $\bullet$ makes $\E$ a formally real
Jordan algebra, with identity element $u$.
\end{itemize} \end{theorem}

In \cite{JvNW}, Jordan, von Neumann and Wigner classified Euclidean Jordan algebras as belonging to one of two broad types, plus one exceptional example. These are 
\begin{itemize} 
\item[(a)] {\bf \emph{Hermitian parts of matrix algebras}} over $\R, \Cx$ or $\Qt$, ordered as usual;
\item[(b)] {\bf \emph{Spin factors}}, in which the normalized state space is a ball of dimension $n$; and 
\item[(c)] {\bf \emph{The Exceptional Jordan Algebra}} of positive $2 \times 2$ hermitian matrices over the Octonions. 
\end{itemize} 
Thus, it would seem that if we can motivate both homogeneity and
self-duality in operational terms, we will go a great way towards
obtaining an operational characterization of finite-dimensional
QM. This problem is taken up in the next section.  We then discuss the
consequences of assuming that a monoidal probabilistic theory
consisting of Jordan models has locally tomographic composites. Here a
theorem of H. Hanche-Olsen \cite{Hanche-Olsen} can be
invoked to show that, so long as the theory contains even a single
instance of the simplest quantum-mechanical system --- a qubit ---
every system allowed by the theory must be the theory must be quantum.

\subsection{Motivating Homogeneity and Self-Duality}

Let us call a model $A$ {\em HSD} (Homogeneous and self-dual) iff its linear hull $\E(A)$ --- or, equivalently, 
its dual, $\V(A)$ --- is homogeneous and self-dual. Why should this be the case? In this section, we discuss 
several possible answers.\\

\noindent{\bf Homogeneity} A model $A$ is {\em uniform} iff the state
space $\Omega$ contains a {\em uniform state} $\mu$, i.e., one taking
constant values $1/n$ on all outcomes of $X(A)$. Of course, this
implies that all tests in $\A(A)$ have cardinality $n$.  For uniform
systems, homogeneity of $\E(A)$ has a straightforward, natural and
physically reasonable interpretation: it asserts that every
non-singular state should be preparable, by means of a
probabilistically reversible transformation, from the uniformly (or
maximally) mixed state. \footnote{One might raise the {\em aesthetic} objection that
  it is awkward to make special reference to the interior state. But
  it is difficult to see how this is any worse aesthetically than
  making special reference to, say, pure states.}. As noted above, homogeneity is also 
  implied by either of the following conditions:
\begin{itemize} 
\item[(a)] Every interior state is the marginal of an isomorphism state 
\item[(b)] Every state is the marginal of a steering state.
\end{itemize} 
Yet another way of arriving at the homogeneity of $\V(A)$ can be found in \cite{Wilce12}. \\

\noindent{\bf Self-Duality} Self-duality seems less clear-cut, but can be obtained as a consequence of certain symmetry assumptions. Perhaps the simplest and most dramatic is the following beautiful result due to M. Mueller and C. Ududec. Call two states $\alpha, \beta \in \Omega(A)$ {\em sharply distinguishable by effects} iff there exists an effect $a$ such that $\alpha(a) = 1$ and $\beta(a) = 0$. Mueller and Ududec call a system {\em bit-symmetric} iff 
every such pair of states can be mapped to any other such pair by a symmetry of the state cone, 
that is, an affine symmetry of $\Omega$. They then prove:

\begin{theorem}[\cite{Mueller-Ududec}] If $\Omega(A)$ is bit-symmetric, then $\V(A)$ (and hence, $\E(A)$) is self-dual. 
\end{theorem} 

It is worth noting that not every self-dual model is bit-symmetric. For instance, if $\Omega$ is a 2-dimensional 
regular $2n+1$-gon, then $\V(\Omega)$ is self-dual, but $\Omega$ is not bit-symmetric. Bit-symmetry is thus a very 
restrictive, yet very plausible, and operationally meaningful, constraint.  

A more involved condition having a somewhat similar flavor, but dealing with the test space structure $X(A)$ rather than the 
pure states of $A$, is worth mentioning. Call $A$ {\em bi-symmetric} iff it is $2$-symmetric under $G(A)$ and if $G(A)$ acts transitively on pure states. As disussed in Section 2.2, it is quite easy to construct such models one at a time.  Recall that $A$ is {\em sharp} iff for every outcome $x$, there is a unique state $\alpha$ with 
$\alpha(x) = 1$. 

\begin{theorem}[\cite{Wilce11}] Let $\C$ be a monoidal probabilistic theory in which every model is bi-symmetric. 
If $A \in \C$ is irreducible and sharp, then $\E(A)$ is self-dual.  \end{theorem}

Another way of obtaining self-duality from bi-symmetry involves the notion of a conjugate system:

\begin{definition}A {\em conjugate} for a model $A$ is a structure $(\overline{A}, \gamma_A, \eta_A)$, where 
$\overline{A}$ is a model, $\gamma_A : A \rightarrow \overline{A}$ is an isomorphism, and 
$\eta_A$ is a bipartite state (on some non-signaling composite) $A \overline{A}$ such that 
\[\eta_{A}(x,\gamma_{A}(x)) = 1/n\]
for every $x \in X(A)$.  We'll call $\gamma_A$ the {\em conjugation
  map} and $\eta_{A}$, the {\em correlator} for the given
conjugate.\end{definition}

\begin{example} {\em Let $A = A(\H)$ be the quantum model associated with a complex Hilbert space $\H$, 
and $\overline{A} = A(\bar{\H})$ associated with the conjugate Hilbert space. Define a mapping $\gamma_{A} : X(\H) \rightarrow X(\bar{\H})$ by $\gamma_{A} : x \mapsto \bar{x}$ 
(strictly speaking, the identity map!). Then, as discussed in Section 3.3, $\eta_{A}(x,\gamma_A(y)) = |\langle \Psi, x \otimes y \rangle|^2 
= \Tr(P_{\Psi}P_{x \otimes y})$ is a correlator. } \end{example}

If $A$ has a conjugate, then it has a conjugate for which the correlator $\eta_A$ is symmetric, 
in the sense that $\eta(x,\gamma_A(y)) = \eta(y, \gamma_A(x))$, and invariant, in the sense that $\eta_{A}(gx, \gamma_{A}(g y)) = \eta(x,\gamma_{A}(y))$. 
Indeed, $\eta^{T}(x,\gamma_{A}(y)) := \eta(y, \gamma_{A}(x))$ is again a correlator; averaging $\eta$ and $\eta^{T}$ 
gives us a symmetric correlator. If $\eta$ is symmetric, then for all symmetries $g \in G(A)$, $\eta^{g}(x,y) = \eta(gx, gy)$ is again a symmetric correlator; averaging over $G$ yields an invariant symmetric correlator. Henceforth, we assume that correlators are symmetric and invariant. It follows that the bilinear form 
\[\Bi(a,b) := \eta(a, \gamma_{A}(b))\] 
is {\em orthogonalizing}, meaning that $\Bi(x,y) = 0$ for all $x \perp y$ in $X(A)$. For the following, see \cite{Wilce11}:

\begin{theorem} Let $A$ be irreducible, bi-symmetric, and have a conjugate $(\bar{A}, \gamma_A, \eta_A)$. Then (a) $\Bi$ is an inner product on $\E$, and (b) $A$ is self-dual with respect to $\Bi$ iff $\eta_{A}$ is an isomorphism state iff $A$ is sharp. \end{theorem}


\subsection{HSD and Jordan Models} 

Call a model $A$ {\em HSD} (Homogeneous and self-dual)
iff the cone $\E_+$ is homogeneous under {\em some} group $\G(A)$ of
order-automorphisms, 
and self-dual with respect to {\em some} inner product.  If $A$ is an HSD model, then by
the Koecher-Vinberg theorem, $\E(A)$ carries a unique euclidean Jordan
structure with respect to which the order unit, $u$, is the
identity and $\langle a, u \rangle = \Tr(a)$. 

An {\em idempotent} in a Jordan algebra $\E$ is an element $e \in \E_+$ with $e^2 = e \bullet e = e$. 
Idempotents in the special Jordan algebra $\L_h(\H)$ are precisely orthogonal projection operators. 
A {\em primitive} idempotent is an idempotent that is not a sum of other non-zero idempotents; thus, in the context 
of $\L_h(\H)$, a primitive idempotent is a rank-one projection operator. Any Euclidean Jordan algebra $\E$ carries 
a canonical trace functional, with $\Tr(ab) = \langle a, b \rangle$, and one can show that $\Tr(e) = 1$ for any 
primitive idempotent. A {\em Jordan frame} in a Euclidean Jordan algebra $\E$ is a set$e_1,...,e_n$ of primitive idempotents summing to $u$. The Spectral Theorem for Euclidean Jordan algebras asserts that every $a \in \E$ has 
a unique representation as a sum of the form $\sum_{e \in E} t_{e} e$ over a Jordan frame $E$, where $\{t_{e} | e \in E\}$ are non-negative real coefficients. 
It follows that the extremal elements of the cone 
$\E_+$ are exactly the primitive idempotents. The group of order-automorphisms of $\E$ fixing  the unit $u$ acts transitively on the set of Jordan frames, so all Jordan frames have the same size, the {\em rank} of $\E$.  (Indeed, regarding the set of 
Jordan frames as a test space, this group acts fully transtively, i.e., any permutation of a Jordan frame can be implemented 
by an order-automorphism of $\E$.)

\begin{definition} A probabilistic model $A$ is {\em uniform} iff its test have a uniform cardinality $n$, and the uniformly mixed probability weight $\mu(x) \equiv 1/n$ belongs to $\Omega(A)$.\end{definition}

 If $A$ is an HSD model, then every primitive idempotent $e$ in $\E(A)$ defines a pure state,
$\langle e |$, and this is the unique pure state assigning probability
$1$ to the effect corresponding to $e$. By a {\em
Jordan model}, we mean an HSD model $A$ such that every outcome in $X(A)$ is a
primitive idempotent in $\E(A)$, or, equivalently, every test is a Jordan frame. Evidently, such a model is unital,  indeed, sharp, and uniform. 

There is a converse. Suppose $A$ is HSD. By an easy extension of the converse to the Krein-Mil'man theorem, any closed, generating subset of $\V(A)_{+}$ contains every 
a point on every extremal ray of $\V(A)_+$. By our standing assumpton of outcome-closure, the outcome-space $X(A)$ is closed in $\E(A)_+$; by construction, it is also generating. 
Since $\V(A)_+ \simeq \E(A)_+$, every extremal ray of $\E(A)_+$ consists of multiples of an outcome. 
Giving $\E(A)$ its standard Jordan structure, primitive idempotents generate extremal rays of $\E(A)_+$, so 
every primitive idempotent in $\E(A)$ is a positive multiple of an outcome in $X(A)$. 


\begin{lemma} Let $A$ be HSD, and let $\E(A)$ have its canonical Jordan structure.  Then:
\begin{itemize} 
\item[(a)] Every extremal unital outcome $x \in X(A)$ is a primitive idempotent. 
\item[(b)] If $A$ is uniform, then every unital outcome is extremal, hence, a primitive idempotent.
\item[(c)] If $A$ is both unital and uniform, it is a Jordan model.
\end{itemize} \end{lemma}

\noindent{\em Proof:} (a) Let $x \in X(A)$ be extremal. As discussed above, there then exists some $t > 0$ such that $tx =: e$, a primitive idempotent. Now suppose $f$ is a primitive idempotent representing a pure state of $\E$, with $\langle f, x \rangle = 1$. Then 
\[t = t \langle f, x \rangle = \langle f, tx \rangle = \langle f, e \rangle \leq 1,\]
by the Cauchy-Schwarz inequality. Now notice that 
\[t^2 \langle x, x \rangle = \langle e, e \rangle = 1\]
so $\langle x, x \rangle = 1/t^2$. Choosing any $E \in \A(A)$ with $x \in E$, we now have 
\begin{eqnarray*}
1 = \langle e, u \rangle  =  t \langle x, u \rangle & = & t \left ( \langle x, x \rangle + \sum_{y \in E \setminus \{x\}} \langle x, y \rangle \right ) 
\geq  t \langle x, x \rangle =  t/t^2 = 1/t,\end{eqnarray*}
so that $t \geq 1$. Thus, $t = 1$, and $x = e$, a primitive idempotent. 

(b) Let $x = \sum_{i} s_i x_i$ where the $x_i$ are extremal outcomes and $s_i \geq 0$. 
Let $\mu$ be the uniform state on $\E$. Then 
\[\frac{1}{m} = \mu(x) = \sum_{i} s_i \mu(x_i) = \sum_i s_i \frac{1}{m} \]
so $\sum_i s_i = 1$. If $x$ is unital, therefore, there exists a primitive idempotent $f$ with 
\[1 = \langle f, x \rangle = \sum_{i} s_i \langle f, x_i\rangle.\]
Since the coefficients $s_i$ are convex, we have $\langle f, x_i \rangle = 1$ for every $i$ with $s_i \not = 0$. 
But then, every $x_i$ is a unital extremal outcome and so, by part (a), a primitive idempotent. It follows 
(again by the Cauchy-Schwarz inequality) that $s_i \not = 0$ implies $x_i = f$, whence, $x = f$ is again a primitive idempotent. (c) now follows at once from (a) and (b). $\Box$\\

\subsection{Composites of Jordan Models} 

Suppose a probabilistic theory $\C$ consists entirely of Jordan
models. Under what conditions can one equip $\C$ with an associative
compositional structure so as to obtain a {\em monoidal} probabilistic
theory? Subject to two further requirements, this this is possible
{\em only} if $\C$ is in fact a standard quantum theory:

\begin{theorem}[\cite{BW12}] Let $\C$ be a symmetric monoidal category of Jordan probabilistic models such that (i) for every $A, B \in \C$, the composite $AB$ is locally tomographic, 
and (ii) at least one system in $\C$ has the structure of a
qubit. Then every model in $\C$ is the hermitian part of a complex
matrix algebra.\end{theorem}

The proof of this result exploits the following theorem due to H. Hanche-Olsen.

\begin{theorem}[Hanche-Olsen] If $\E$ is a JC (check) algebra and $\M_2$ is the Jordan algebra of $2 \times 2$ hermitian matrices over $\Cx$, then $\E$ is the Hermitian part of a complex matrix algebra iff there exists a Jordan product on $\E \otimes \M_2$ such that
\begin{equation} (a \otimes \1)\bullet (b \otimes \1) = ab \otimes \1 \ \mbox{and} \ (\1 \otimes x) \bullet (\1 \otimes y) = \1 \otimes xy\end{equation} 
for all $a, b \in \E$ and all $x,y \in \M_2$.\end{theorem}

Essentially, \cite{BW12} shows that if $AB$ is a non-signaling HSD composite of HSD models $A$ and $B$, then 
local tomography forces the Jordan product on $\E(AB)$ to satisfy (\theequation). A key step is the following observation. 

\begin{lemma} Suppose $A$ is a Jordan model. Let $AA$ be a non-signaling composite of $A$ with itself. If $AA$ is Jordan, 
then the trace form on $\E(AA)$ factors. \end{lemma}

\noindent \noindent{\em Proof:} By definition of a composite, if $x, y \in X(A)$, then $x
\otimes y$ is an outcome in $X(AA)$. Since $x$ and $y$ are unital in 
$A$, $x \otimes y$ is unital in $X(AA)$. Indeed, the pure product
state $\langle x | \otimes \langle y|$ assigns $x \otimes y$
probability $1$ (again, by definition of a composite). Hence, by Lemma
part (b) of Lemma 8, $x \otimes y$ is a primitive idempotent in $\E(AA)$. But then
we also have $\langle x \otimes y | x \otimes y \rangle = 1$, and this
is the unique pure state with this property.  Hence, $\langle x |
\otimes \langle y | = \langle x \otimes y |$, so that
\[\langle x \otimes y | a \otimes b \rangle = \langle x | a \rangle \langle y | b \rangle\]
for all $a, b \in \E(A)$. Since $X(A)$ spans $\E(A)$, the same holds with arbitrary elements of $\E(A)$ in place of 
$x$ and $y$, i.e, the inner product factors. $\Box$\\ 

Local tomography is a strong constraint on a probabilistic theory. The
fact that real and quaternionic quantum mechanics are not locally
tomographic should at least slightly temper our willingness to adopt
it. A classification of non-locally tomographic non-signaling
composites of Jordan models is the subject of on-going work.

\section{Conclusion}

The framework we have sketched here for a post-classical probability
theory has several virtues. It is conceptually conservative,
mathematically straightforward, and easily accommodates free
mathematical constructions, as well as the introduction of further
structure (for example, one can readily topologize the concept of a
test space; see \cite{Wilce05a, Wilce05b}). Still, at present, what we
have is indeed just the sketch of a framework. Its further development
offers many interesting opportunities. We close by mentioning five
areas for further work.\\

\noindent{\em Quantum Axiomatics.} As long as we restrict our
attention to finite-dimensional probabilistic models, it seems that
there are many different axiomatic packages --- that is, many
different clusters of plausible constraints --- that locate orthodox
QM, or its near environs, within the wild landscape of general
post-classical probabilistic theories. In addition to the approach via
homogeneity and self-duality, sketched in Section 4, there are various
derivations of finite-dimensional QM in the spirit of Hardy's axioms
\cite{Hardy}, including work by Rau \cite{Rau}, Dakic and Brukner
\cite{Dakic-Brukner}, Masanes and Mueller \cite{Masanes-Mueller} and
Chiribella, D'Ariano and Perinotti \cite{CDP}. A different approach
\cite{Goyal} exploits information geometry. There is also the
completeness theorem of Selinger \cite{SelingerComplete} for
dagger-compact categories. This is not even to mention the various
axiomatic treatments of quantum theory given in the older
quantum-logical literature. (This last has been criticized as being too
``mathematical", but much of it becomes significantly simpler when specialized
to the finite-dimensional case.) It would be of great interest to know
how all of these various axiomatizations (most of which share at least
a few assumptions), are related to one another. The mathematical
framework developed here seems ideal for this task.\\

\noindent{\em Infinite-Dimensional Models} Of even greater interest
would be to extend the results of these efforts to
infinite-dimensional settings. Individually, infinite-dimensional
probabilistic models have been well-studied \cite{Davies-Lewis,
  Edwards}, and tools are available for dealing with composites in
this setting, too \cite{Wilce92}. However, the line of argument
developed in Section 5, depending as it does on the Koecher-Vinberg
Theorem, does not generalize easily to the infinite-dimensional
setting. Efforts in this direction are just getting underway [refs?],
but there is a great deal more work to be done. \\

\noindent{\em Quantum Field Theory} Algebraic quantum field theory
associates an algebra of observables to each open subset of
spacetime. An obvious project would be to consider a probabilistic
theory in which each such region is associated with a probabilistic
model, subject to the constraint that the model associated with a
union of spacelike separated regions be a non-signaling composite of
the models associated with the regions individually.  \\ 

\noindent{\em Applications; Post-Quantum Information Theory} The
notion of a probabilistic model is very broad. It would likely be a
fruitful exercise to look for applications outside of quantum
information and the foundations of quantum mechanics in which models
that are neither classical nor quantum arise. In anticipation of this,
it would be very reasonable to further develop the post-classical
information theory sketched in \cite{Entropy, SWentropy}, especially
by investigating in some detail such ideas as {\em channel capacity} in
this setting. \\

\noindent{\em The Measurement Problem.} Even though we take
measurements and measurement-outcomes as primitives, nothing prevents
us from asking whether these can be modeled dynamically {\em within}
the formal framework presented here. Certain versions of the
measurement problem can be formulated as theorems in this framework,
leading one to wonder whether various strategies for resolving the
{\em quantum} measurement problem --- e.g., some version of ``many
worlds" interpretations, or the apparatus of decoherence --- have analogues 
 in the setting of a general probabilistic
theory. If so, this would shed some light on {\em how} these
interpretive moves work; if not, then the existence  of such an analogue 
could be regarded as another constraint on a probabilistic theory,
taking us closer to orthodox QM. A further discussion of these matters
can be found in \cite{Wilce10}.  


\end{document}